%% file: main.tex
\documentclass[aps, pra, amsmath, amssymb, 11pt, final, tightenlines, twoside, twocolumn, nofloats, nofootinbib, superscriptaddress, showkeys, showkeywords]{revtex4-2}

\usepackage[T1]{fontenc}
\usepackage{lmodern}
\usepackage[utf8]{inputenc}
\usepackage[english]{babel}
\usepackage{graphicx}
\graphicspath{{figures/}}
\usepackage{dcolumn}
\usepackage{bm}

\newcommand{\Ms}{$\textrm{M}_{\odot}$}

\newcommand{\oi}{\,{\sc i}}
\newcommand{\ii}{\,{\sc ii}}
\newcommand{\iii}{\,{\sc iii}}
\newcommand{\Ha}{\ensuremath{\mathrm{H}\alpha}}
\newcommand{\Hb}{H$\beta$}
\newcommand{\Hg}{H$\gamma$}

\newcommand{\nb}{\textsc{NBursts}}

\input{maik.rty}

\setcitestyle{authoryear,round}
\input{sao_cmd_author.tex}

\providecommand{\texorpdfstring}[2]{#1}

\begin{document}

\selectlanguage{english}

\keywords{galaxies: individual: SDSS J113356.96+511459.2, SDSS J105252.58+432542.2; galaxies: stellar populations; galaxies: kinematics and dynamics; galaxies: elliptical and lenticular. Astrophysics - Astrophysics of Galaxies}


\title{STELLAR POPULATION AND DYNAMICAL MODELING OF GALAXIES WITH KINEMATICALLY MISALIGNED COMPONENTS: AGE AND METALLICITY OF STRUCTURAL COMPONENTS}

\author{\firstname{V.~S.}~\surname{Goradzhanov}}
 \email{vgoradzanov@gmail.com}
 \affiliation{Sternberg Astronomical Institute, Moscow, 119234, Russia}
 \affiliation{Department of Physics, M.~V.~Lomonosov Moscow State University, Moscow, 119991}

\author{\firstname{K.~O.}~\surname{Parfenov}}
 \affiliation{Astrospace Center of the P.~N.~Lebedev Physical Institute, Russian Academy of Sciences, Moscow, Russia}

\author{\firstname{D.~D.}~\surname{Mishurin}}
 \affiliation{School No.\ 2126 ``Perovo'', Moscow, Russia}

\author{\firstname{O.~K.}~\surname{Sil'chenko}}
 \affiliation{Sternberg Astronomical Institute, Moscow, 119234, Russia}
 \affiliation{Department of Physics, M.~V.~Lomonosov Moscow State University, Moscow, 119991}

\begin{abstract}
We present the results of stellar population and orbit modeling of two galaxies with kinematically misaligned components---PGC~35706 and LEDA~2220522---extending their Schwarzschild dynamical modeling \citep{Goradzhanov2025}. Using spatially resolved stellar population age and metallicity maps obtained by \textsc{NBursts} full spectral fitting with the \textsc{E-MILES} grid for \textsc{MaNGA} survey data, we solve a regularized linear inverse problem with bounds (\textsc{BVLS}) that assigns to each orbital cell in the $(R,\lambda_z)$ plane its own light-weighted $T_\mathrm{SSP}$, $[\mathrm{M/H}]_\mathrm{SSP}$, and $M/L$ values. Mean component parameters and their radial profiles are calculated using projected mass within the actual \textsc{MaNGA} field of view; profile segments with a local component fraction below $1\%$ are marked as weakly constrained. In LEDA~2220522, the counter-rotating disk is younger than the co-rotating disk: $1.32\pm0.04$ versus $1.61\pm0.06$~Gyr. It is also more metal-rich: $[\mathrm{M/H}]=-0.450\pm0.043$ versus $-0.613\pm0.024$. The agreement between the rotation directions of the \Ha\ ionized gas and the young counter-rotating component supports a scenario involving the late supply of chemically enriched retrograde gas, but does not require multiple accretion episodes with alternating angular momentum directions. In PGC~35706, the counter-rotating disk contains $8.8\%$ of the mass in the field and is formally older than the co-rotating disk by $1.36\pm0.51$~Gyr and more metal-poor by $0.130\pm0.046$~dex. This combination differs from the commonly observed picture of a young secondary component; however, these differences are regarded as preliminary because of the limited support from the projected data. An analysis of 50 Monte Carlo realizations confirms the stability of the projected maps against statistical noise, but not the uniqueness of the values in individual phase-space cells. The results show that the stellar population and orbit approach is applicable to \textsc{MaNGA} data when the local projected component fraction and systematic effects of regularization and the dynamical model are taken into account.
\end{abstract}

\maketitle

\input{body.tex}

\bibliographystyle{aspb1}
\bibliography{Article}

\end{document}

%% file: sao_cmd_author.tex
\def\squareforqed{\hbox{\rlap{$\sqcap$}$\sqcup$}}

\def\sq{\ifmmode\squareforqed\else{\unskip\nobreak\hfil
\penalty50\hskip1em\null\nobreak\hfil\squareforqed
\parfillskip=0pt\finalhyphendemerits=0\endgraf}\fi}

\def\arcsec{\hbox{$^{\prime\prime}$}}

\def\utw{\smash{\rlap{\lower5pt\hbox{$\sim$}}}}

\def\udtw{\smash{\rlap{\lower6pt\hbox{$\approx$}}}}

\def\diameter{{\ifmmode\mathchoice
{\ooalign{\hfil\hbox{$\displaystyle/$}\hfil\crcr
{\hbox{$\displaystyle\mathchar"20D$}}}}
{\ooalign{\hfil\hbox{$\textstyle/$}\hfil\crcr
{\hbox{$\textstyle\mathchar"20D$}}}}
{\ooalign{\hfil\hbox{$\scriptstyle/$}\hfil\crcr
{\hbox{$\scriptstyle\mathchar"20D$}}}}
{\ooalign{\hfil\hbox{$\scriptscriptstyle/$}\hfil\crcr
{\hbox{$\scriptscriptstyle\mathchar"20D$}}}}
\else{\ooalign{\hfil/\hfil\crcr\mathhexbox20D}}%
\fi}}

%% file: body.tex
%
%

\section{INTRODUCTION}
\label{sec:intro}

The study of galaxies with kinematically misaligned components---systems in which rotating stellar and gaseous subsystems with different rotation directions or mutually misaligned geometries coexist---provides one of the most direct observational ways to trace the history of accretion processes and interactions in galaxy evolution \citep{Katkov2013, Katkov2016, Corsini2014}. The most striking manifestation of such configurations is counter-rotating stellar disks, first reliably detected in NGC~4550 \citep{Rix1992, Rubin1992} and since then systematically identified in integral-field spectroscopic data both for individual objects \citep{Coccato2011, Coccato2013, Johnston2013, Mitzkus2017, Morelli2017, Katkov2024} and in large samples \citep{Bao2022, Bevacqua2022, Gasymov2025}. The very presence of two disks with opposite rotation directions rules out purely internal formation scenarios and requires external processes: accretion of gas with opposite angular momentum from the intergalactic medium \citep{Sancisi2008, Bournaud2005}, minor mergers with gas-rich satellites \citep{Thakar1996, Bassett2017}, or combinations of these processes \citep{Algorry2014, Khoperskov2021, Starkenburg2019}.

Kinematics, however, specifies only the geometry of the present state of a system and cannot by itself reconstruct the sequence of events that led to the formation of its multicomponent structure. A key additional source of information is provided by stellar population properties, primarily the light-weighted SSP-equivalent age $T_\mathrm{SSP}$ and metallicity $[\mathrm{M/H}]_\mathrm{SSP}$ of the two subsystems. Differences in these properties constrain the relative sequence of component formation and the degree of chemical processing of the original material, but do not directly date the dynamical accretion event. In the best-studied systems (NGC~3593, NGC~4550, NGC~5719, NGC~4191, NGC~1366, and several others), the counter-rotating component is generally younger and more metal-poor than the mass-dominant disk \citep{Coccato2011, Coccato2013, Johnston2013, Coccato2015, Morelli2017}. This is interpreted as a consequence of its stars forming from material that has not undergone chemical processing in the original galaxy. For more massive or ``mature'' counter-rotating disks in the \textsc{MaNGA} and \textsc{SAMI} samples, age differences are less pronounced, consistent with earlier formation or prolonged mixing of the populations \citep{Bevacqua2022, Bao2022}. Additional diagnostic information could be provided by $\alpha$-element abundances, which constrain the duration of the star formation epoch \citep{Thomas2005, delaRosa2011}; however, their reliable extraction from maps with the spectral resolution and signal-to-noise ratio typical of \textsc{MaNGA} is difficult and is not attempted in this work.

Spatially resolved age and $[\mathrm{M/H}]$ maps derived from full spectral fitting are likewise insufficient on their own for a rigorous interpretation: each pixel in the observed field of view integrates light from stars that generally belong to different dynamical subsystems. A natural step is to combine spatially resolved stellar population maps with an orbitally decomposed galaxy model. This approach, proposed by \citet{Poci2019} and validated using mock data by \citet{Zhu2020}, uses the Schwarzschild orbit-superposition method \citep{Schwarzschild1979, 2008_triaxial_van_den_Bosch, Zhu2018a, Zhu2018b} to define orbital ``cells'' in the $(R,\lambda_z)$ plane and then assigns a mean age and metallicity to each cell by solving a linear inverse problem using two-dimensional $T_\mathrm{SSP}$ and $[\mathrm{M/H}]_\mathrm{SSP}$ maps. The method has been successfully applied to high-resolution \textsc{MUSE} data for the lenticular galaxy NGC~3115 \citep{Poci2019}, a sample of S0 galaxies in the Fornax cluster within the \textsc{Fornax3D} project \citep{Poci2021, Zhu2022a, Zhu2022b, Ding2023}, and---at substantially lower spatial resolution---spiral galaxies from the \textsc{CALIFA} survey \citep{Jin2024}. In contrast, applying the stellar population and orbit approach to \textsc{MaNGA} data, which provide a large, statistically representative sample of galaxies with kinematically misaligned components, remains largely unexplored, both because of the lower spatial resolution compared to \textsc{MUSE} and because several kinematically similar components must be separated within a small field of view.

Previously, we presented dynamical models of two galaxies with kinematically misaligned components---PGC~35706 and LEDA~2220522---constructed using the Schwarzschild orbit-superposition method in the axisymmetric approximation, based on \textsc{MaNGA} integral-field spectroscopy and deep \textsc{DESI} photometry \citep{Goradzhanov2025}. In both galaxies, three structural components were identified using the orbital circularity parameter $\lambda_z$: a spheroidal component (bulge/halo), a co-rotating disk, and a counter-rotating disk. An observation central to the subsequent interpretation is the agreement between the rotation directions of the ionized gas (traced by \Ha) and the counter-rotating stellar disk in both systems; this links the gas to the retrograde component and supports an external supply of material. However, the relative ages and chemical properties of the identified stellar components were not addressed: a purely dynamical model cannot establish whether their stellar populations differ and whether these differences are consistent with late formation of the secondary disk.

The present work is a natural continuation of \citet{Goradzhanov2025} and is devoted to stellar population and orbit modeling of the same two galaxies using the methodology proposed by \citet{Poci2019} and developed by \citet{Zhu2020}. To make the presentation self-contained, we construct the dynamical model of each galaxy anew using the Schwarzschild orbit-superposition method, fixing the dark halo core radius at the effective radius $R_e$ and fitting its normalization (Section~\ref{subsec:data_halo}). We combine the reconstructed orbital structure with spatially resolved age, metallicity, and $M/L$ maps obtained by \nb\ full spectral fitting \citep{nburst_a, nburst_b} with the \textsc{E-MILES} model grid \citep{Vazdekis2016MNRAS.463.3409V}, solving a regularized linear inverse problem with bounds for orbital cells in the $(R,\lambda_z)$ plane. The main aim is to compare the stellar populations of the co-rotating and counter-rotating disks in two specific galaxies and to test whether their relative ages and metallicities are consistent with the late supply of retrograde material. Additional objectives are:
\begin{enumerate}
\item to construct radial age and $[\mathrm{M/H}]$ profiles for all three structural components using projected mass within the \textsc{MaNGA} field and to assess their sensitivity to statistical noise using Monte Carlo simulations;
\item to compare the inferred differences qualitatively with the expected signatures of gas accretion and minor mergers, without attempting to select a unique formation channel on the basis of two objects \citep{Algorry2014, Bassett2017, Starkenburg2019, Khoperskov2021};
\item to assess how the interpretability of component parameters depends on the local projected component fraction and the limited spatial coverage of \textsc{MaNGA}.
\end{enumerate}
Because the sample contains only two galaxies, the results are treated as detailed studies of individual systems, rather than as a statistical test of the frequency or universality of any formation scenario.

The paper is organized as follows. Section~\ref{sec:data} briefly describes the observational data and input kinematic and stellar population maps. Section~\ref{sec:method} presents the stellar population and orbit modeling methodology. Section~\ref{sec:results} gives the main results: the reconstructed component ages and metallicities in PGC~35706 and LEDA~2220522. Their interpretation in the context of formation scenarios is discussed in Section~\ref{sec:discussion}, and the main conclusions are summarized in Section~\ref{sec:conclusions}.

\section{DATA}
\label{sec:data}

This work uses the same observational data as \citet{Goradzhanov2025}, to which we refer the reader for a more detailed description of their reduction. Here we provide only a brief summary and focus on the aspects relevant to stellar population and orbit modeling, primarily the spatially resolved stellar population age and metallicity maps.

\subsection{\textsc{MaNGA} integral-field spectroscopy
and \textsc{DESI} photometry}
\label{subsec:data_manga_desi}

\begin{figure*}[t!]
\begin{minipage}[h]{0.49\linewidth}
\center{\includegraphics[trim=0cm 0cm 0cm 0cm, clip, width=\linewidth]{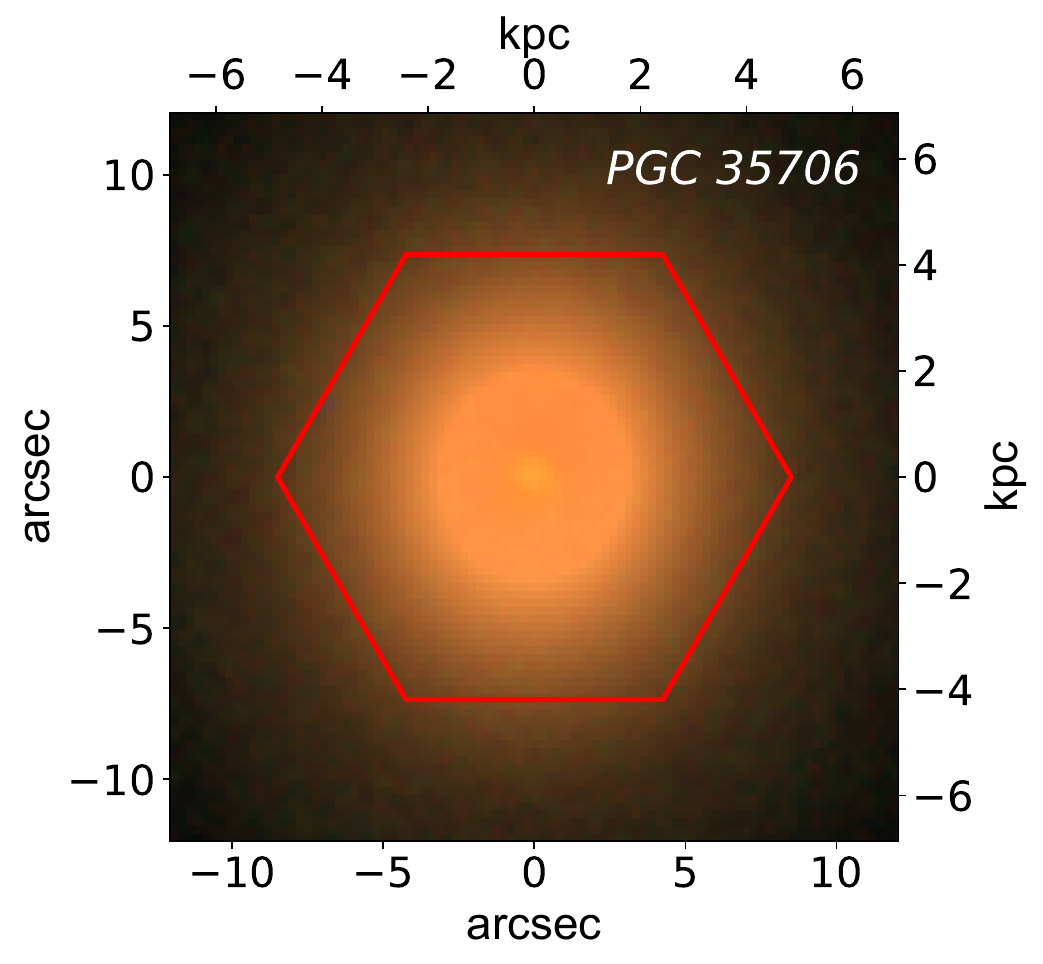}}
\end{minipage}
\hfill
\begin{minipage}[h]{0.49\linewidth}
\includegraphics[trim=0cm 0cm 0cm 0cm, clip, width=\textwidth]{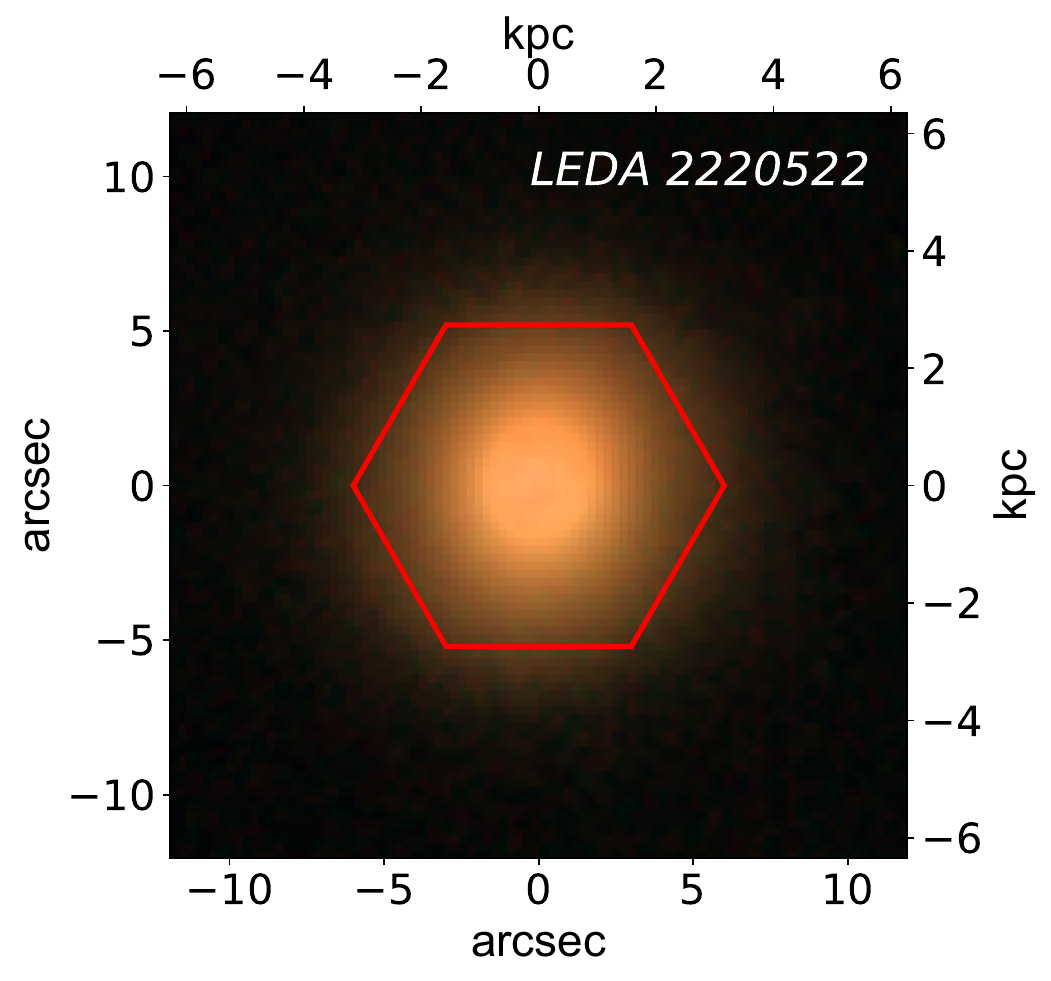}
\end{minipage}
\caption{Photometric images of PGC~35706 and
LEDA~2220522: illustrative RGB composites of $g$-, $r$-, and $z$-band images from the \textsc{DESI} survey. The red hexagon marks the \textsc{MaNGA} field of view. The photometric \textsc{MGE} decomposition uses separate monochromatic $z$-band images rather than these color composites.}
\label{ris:DESI_images}
\end{figure*}

Both galaxies considered here---PGC~35706 (SDSS~J113356.96+511459.2) and LEDA~2220522 (SDSS~J105252.58+432542.2)---were observed as part of the final \textsc{MaNGA} survey release (\textsc{DR17}; \citealt{Abdurrouf2022, Blanton2017}) with hexagonal fiber bundles of different sizes: a 37-fiber bundle (field of view $\sim\!17.5\arcsec$) for PGC~35706 and a 19-fiber bundle ($\sim\!12.5\arcsec$) for LEDA~2220522, using the 2.5-m Sloan Foundation Telescope and the \textsc{BOSS} spectrographs \citep{Smee2013}. The initial reduction was performed with the standard \textsc{DRP} pipeline \citep{Law2016}; the data are provided as three-dimensional cubes, with a wavelength calibration accuracy of $\sim5~$km/s at a median spectral resolution of $R \sim 2000$ ($\sim 72~$km/s). Deep $z$-band photometry of both galaxies was obtained from the \textsc{DESI} survey \citep{Dey2019} and used by \citet{Goradzhanov2025} to construct surface brightness models using multi-Gaussian expansion (\textsc{MGE}; \citealt{MGE_Capp}) and subsequently to calculate the gravitational potential with the \textsc{AGAMA} library \citep{AGAMA}. The $z$ band was chosen as the longest-wavelength optical band available: its surface brightness is less sensitive to dust extinction \citep{Schlafly2011} and to the local contribution of young blue stars, and therefore better traces the distribution of the old population that accounts for most of the stellar mass. In addition, this is the band for which the \textsc{E-MILES} models provide the mass-to-light ratio $M/L_z$ used below; using the same band for the \textsc{MGE} luminosity and $M/L$ ensures a consistent normalization of the stellar density. The color images in Fig.~\ref{ris:DESI_images} are illustrative RGB composites of the $g$, $r$, and $z$ bands; separate monochromatic $z$-band images alone are used to construct the \textsc{MGE} and the potential. In Fig.~\ref{ris:DESI_images}, LEDA~2220522 does indeed appear somewhat more flattened than PGC~35706. This is not inconsistent with its smaller adopted inclination, because the observed shape also depends on the intrinsic flattening of the galaxy.

The basic object parameters adopted in the modeling are listed in Table~\ref{tab:basic_parameters}. Redshifts and stellar masses are taken from the \textsc{MaNGA DynPop} catalog \citep{ZhuDynPop2023}; distances, angular scales, inclinations, and effective radii correspond to the inputs of the dynamical models actually used. The catalog mass is converted from its original units of $h^{-2}M_\odot$ to physical $M_\odot$ using $h=0.73$, as adopted in the abundance-matching check, and is used only in this external check of the halo normalization, rather than as a constraint in the main orbital fit.

\begin{table}[ht!]
\resizebox{\columnwidth}{!}{%
\begin{tabular}{lcc}
\hline
\hline
Parameter & PGC~35706 & LEDA~2220522 \\
\hline
\textsc{MaNGA} plate--IFU & 8992--3704 & 8254--1902 \\
$z$ & 0.02610 & 0.02424 \\
adopted distance, Mpc & 117.490 & 108.643 \\
scale, kpc/$\arcsec$ & 0.570 & 0.527 \\
inclination $i$ & $42^\circ$ & $27^\circ$ \\
$R_e$, $\arcsec$ & 3.48 & 2.52 \\
$R_e$, kpc & 1.98 & 1.33 \\
$\log(M_\star/M_\odot)$ ($h=0.73$) & 10.713 & 10.014 \\
\hline
\end{tabular}
}
\caption{Basic galaxy parameters. Redshifts and stellar masses are taken from the \textsc{MaNGA DynPop} catalog \citep{ZhuDynPop2023}; catalog masses in units of $h^{-2}M_\odot$ are converted to physical masses using $h=0.73$. The remaining quantities are the adopted input parameters of the dynamical models in this work. The effective radius is obtained from the photometric \textsc{MGE} decomposition used here.}
\label{tab:basic_parameters}
\end{table}

\subsection{Kinematic and stellar population parameter maps}
\label{subsec:data_maps}

The age and metallicity maps were obtained previously by \citet{Goradzhanov2025}: spaxels in the \textsc{MaNGA} cubes were combined into Voronoi bins \citep{Cappellari2003MNRAS.342..345C} with a target signal-to-noise ratio of $\sim 20$, and full spectral fitting was performed in each bin using \nb\ \citep{nburst_a, nburst_b} with the \textsc{E-MILES} grid of simple stellar population (SSP) models \citep{Vazdekis2016MNRAS.463.3409V}, spanning ages $0.1 \lesssim T_\mathrm{SSP} \lesssim 14~$Gyr and metallicities $-1.71 \lesssim [\mathrm{M/H}]_\mathrm{SSP} \lesssim +0.40~\mathrm{dex}$ (\textsc{BaSTI} isochrones). The $\chi^2$ minimization simultaneously determined:
\begin{enumerate}
\item the parameters of the stellar Gauss-Hermite line-of-sight velocity distribution (LOSVD): $V$, $\sigma$, $h_3$, and $h_4$;
\item a separate LOSVD for the bright emission lines (\Hg, \Hb, [O\iii], [O\oi], [N\ii], \Ha, [S\ii]), approximating the ionized-gas kinematics;
\item light-weighted age $T_\mathrm{SSP}$ and metallicity $[\mathrm{M/H}]_\mathrm{SSP}$ values, obtained by interpolation within the \textsc{E-MILES} grid.
\end{enumerate}
To reproduce the observed continuum shape correctly, the model was multiplied by a polynomial of degree $19$, compensating for interstellar extinction and possible errors in the spectral sensitivity calibration.

In the \nb\ configuration used here, each spectrum is described by a single SSP template; thus, the inferred $T_\mathrm{SSP}$ is an SSP-equivalent age rather than a reconstructed star formation history. The fit does not determine the set of mass fractions in age intervals required to calculate $\mathrm{SFR}(t)$; consequently, these SSP models cannot yield star formation rates for the co-rotating and counter-rotating disks or allow their comparison. The \Ha\ emission independently characterizes current star formation in the gas, but without additional spectral decomposition it cannot be uniquely assigned to the two already formed stellar orbital components.

\begin{figure*}[ht!]
\includegraphics[width=\linewidth]{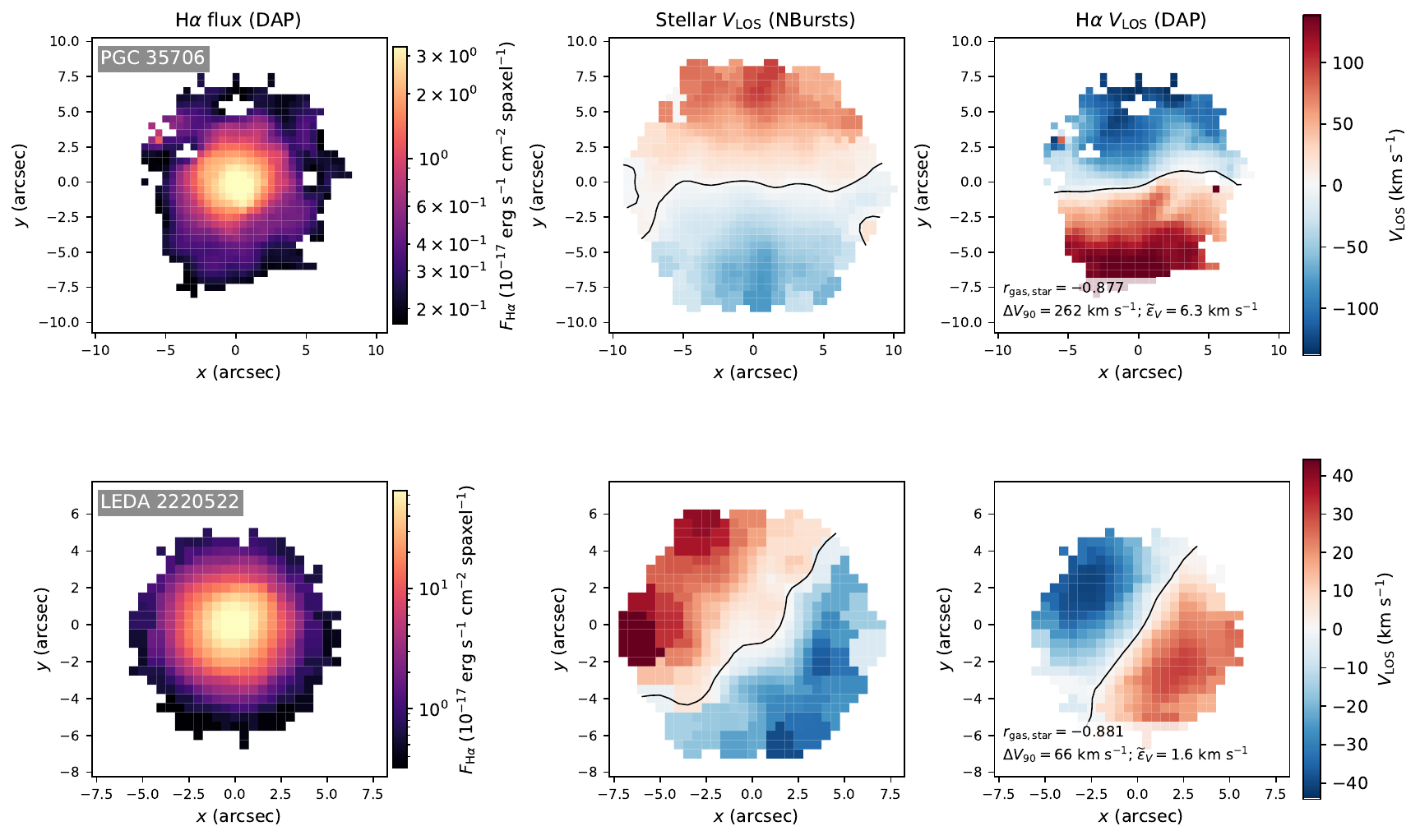}
\caption{Ionized-gas flux and kinematics for PGC~35706 (top row) and LEDA~2220522 (bottom row). Left: \Ha\ line flux from \textsc{MaNGA DAP} in spaxels with $S/N(\mathrm{H}\alpha)\geq5$; center: stellar line-of-sight velocity field from \nb\ fitting; right: \Ha\ line-of-sight velocity field from \textsc{DAP}. Velocities are given relative to the median of each field; black lines show the zero-velocity contours. The negative correlation between the gaseous and stellar fields ($r=-0.877$ and $-0.881$) demonstrates the opposite sense of their large-scale rotation. The range $V_{95}-V_5$ is $262$ and $66$~km/s, with median formal \Ha\ velocity errors of $6.3$ and $1.6$~km/s, respectively.}
\label{ris:Halpha_kinematics}
\end{figure*}

In \citet{Goradzhanov2025}, the stellar and gaseous kinematic maps were obtained in a single spectral fit; however, the \Ha\ velocity field was not included in the orbital optimization and was used only for subsequent comparison with the kinematics of the identified components. In the present work, orbital weights are determined from the stellar density and four stellar kinematic maps ($V$, $\sigma$, $h_3$, $h_4$); the gas kinematics is likewise excluded from the objective function and serves as an independent check. Ionized-gas rotation is securely detected in both galaxies (Fig.~\ref{ris:Halpha_kinematics}). Selecting spaxels with $S/N(\mathrm{H}\alpha)\geq5$ in the \textsc{MaNGA DAP} maps gives a difference between the 95th and 5th percentiles of the line-of-sight velocity field of $262$~km/s for PGC~35706 and $66$~km/s for LEDA~2220522, whereas the median formal errors for individual spaxels are $6.3$ and $1.6$~km/s, respectively. The \textsc{DAP} gas velocity fields are anticorrelated with the stellar fields obtained by \nb\ fitting: the Pearson coefficient is $-0.877$ for PGC~35706 and $-0.881$ for LEDA~2220522. This confirms that the gas has the opposite sense of large-scale rotation to the light-dominant stellar component; the substantial noncircular motions in PGC~35706 are addressed separately below in the discussion of the gas rotation curve.

\begin{figure*}[ht!]
\begin{minipage}[h]{0.49\linewidth}
\center{\includegraphics[trim=0cm 0cm 0cm 0cm, clip, width=1\linewidth]{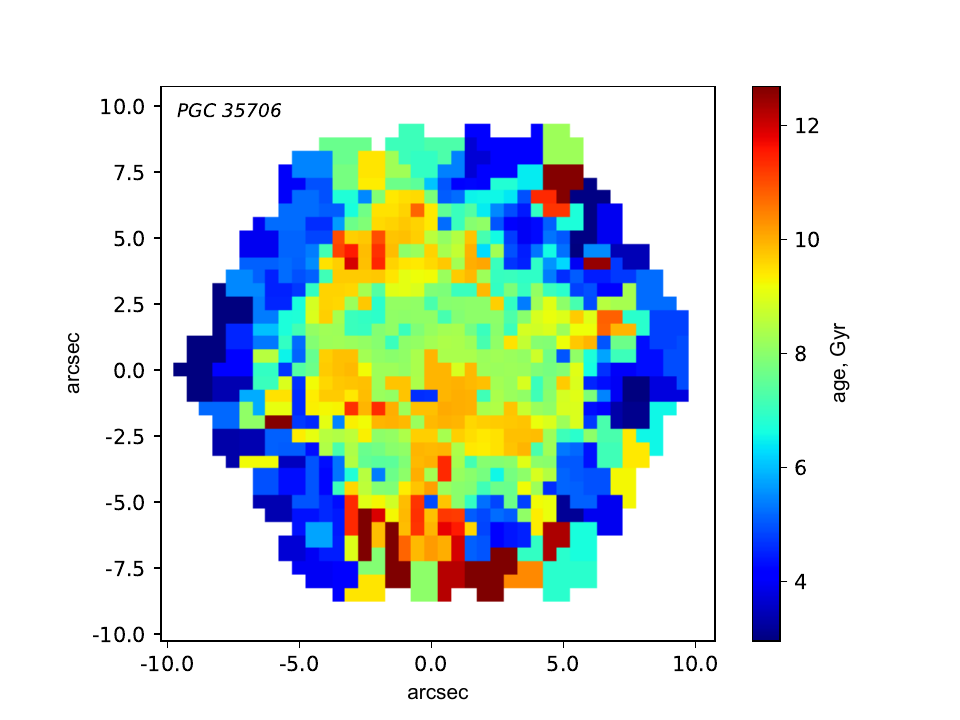}}
\end{minipage}
\hfill
\begin{minipage}[h]{0.49\linewidth}
\center{\includegraphics[trim=0cm 0cm 0cm 0cm, clip, width=1\linewidth]{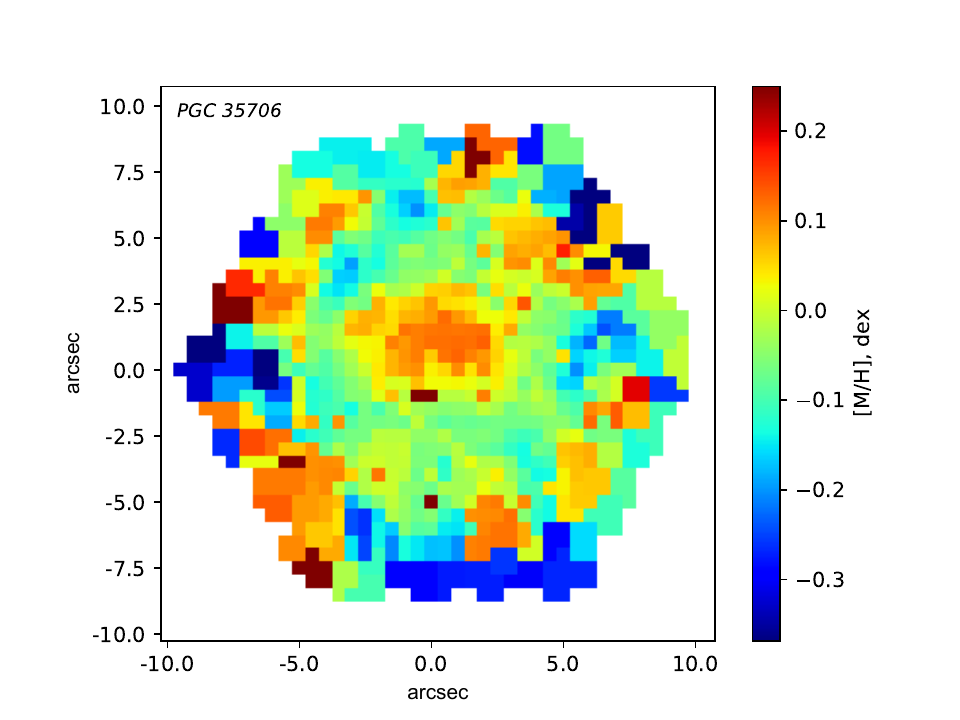}}
\end{minipage}

\begin{minipage}[h]{0.49\linewidth}
\center{\includegraphics[trim=0cm 0cm 0cm 0cm, clip, width=1\linewidth]{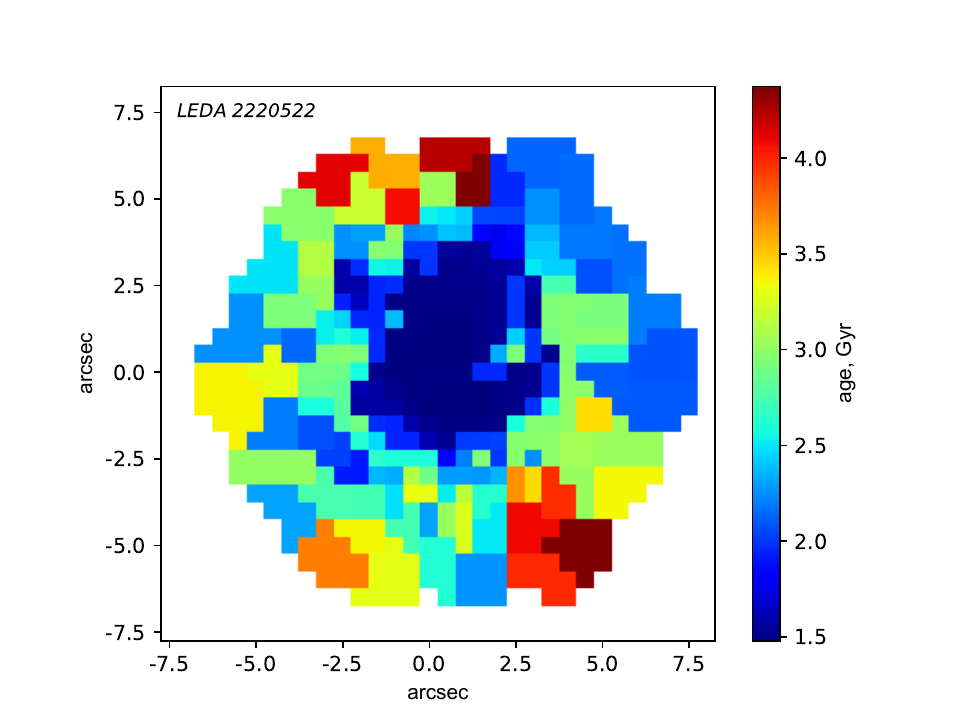}}
\end{minipage}
\hfill
\begin{minipage}[h]{0.49\linewidth}
\center{\includegraphics[trim=0cm 0cm 0cm 0cm, clip, width=1\linewidth]{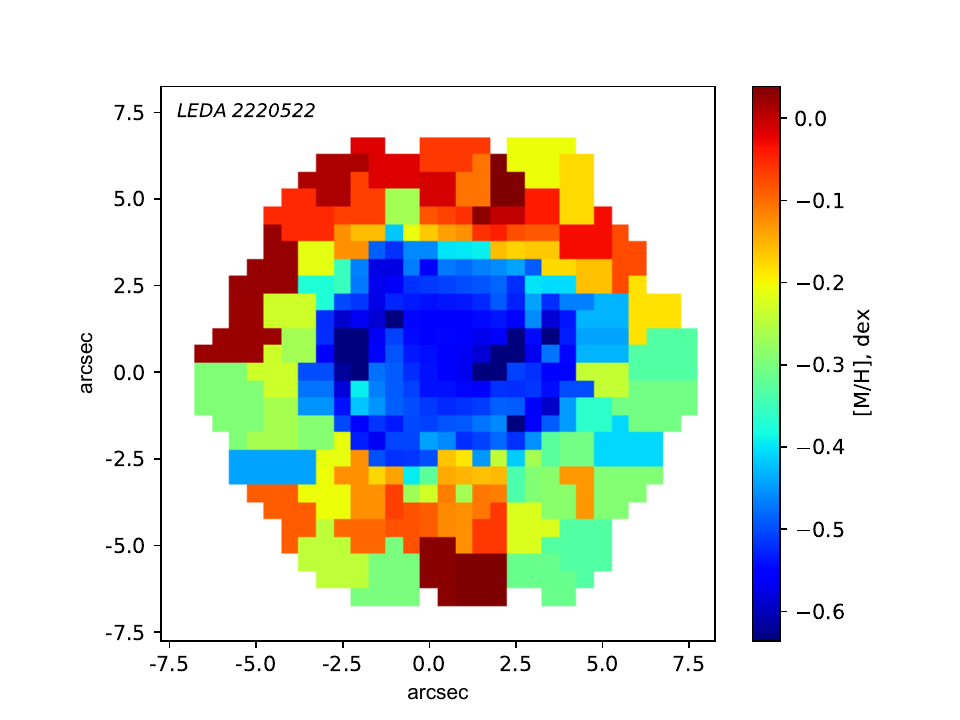}}
\end{minipage}
\caption{Observed two-dimensional maps of light-weighted stellar population
age (left column) and metallicity (right column) for PGC~35706 (top row) and LEDA~2220522 (bottom row), obtained by \nb\ full spectral fitting in Voronoi bins with a target $\mathrm{S/N}\sim20$. The same binning scheme is used for age and metallicity in each galaxy; the apparent difference in region sizes arises because adjacent bins with similar parameter values visually merge on the corresponding color scale. Age is given in Gyr and metallicity in dex. The axes are marked in arcseconds from the photometric center of the galaxy. These maps are the inputs to the stellar population and orbit model (Section~\ref{sec:method}).}
\label{ris:obs_popmaps}
\end{figure*}

In this work, the two-dimensional $T_\mathrm{SSP}$ and $[\mathrm{M/H}]_\mathrm{SSP}$ maps obtained by \citet{Goradzhanov2025} (Fig.~\ref{ris:obs_popmaps}) provide additional inputs; below, we refer to them as the age and metallicity maps for brevity. The maps show clear differences between the two galaxies: PGC~35706 has an older population on average, with a prominent metal-rich nucleus, whereas LEDA~2220522 is substantially younger, with a mean age of $\sim\!1$--$2$~Gyr in the central region and predominantly subsolar metallicity (with a local peak reaching near-solar values at intermediate radii). A detailed quantitative analysis of the radial distributions is presented in Section~\ref{sec:results}.

\subsection{Dynamical model inputs}
\label{subsec:data_dyn_input}

The photometric \textsc{MGE} decomposition was reconstructed after the first paper: in the present model, it contains seven Gaussians per galaxy rather than six. After PSF convolution, the seven-component models reproduce the \textsc{DESI} radial profiles with a median absolute relative residual of about $9\%$; adding an eighth nonnegative Gaussian provides no meaningful improvement. The total luminosity is calculated by analytically integrating all seven components: the circle extending to the maximum radius of valid \textsc{MaNGA} spaxels encloses $90.2\%$ and $94.0\%$ of the integral for PGC~35706 and LEDA~2220522, respectively, while the region within three standard deviations of the broadest Gaussian encloses $99.5\%$ and $99.9\%$. Thus, the extrapolated contribution beyond the maximum observed radius is about $9.8\%$ and $6.0\%$.

In addition to the age and metallicity maps, the stellar population and orbit model takes as input a dynamical model of each galaxy, constructed using the Schwarzschild orbit-superposition method \citep{Schwarzschild1979} as implemented in \textsc{Forstand} \citep{Forstand} within the \textsc{AGAMA} library \citep{AGAMA}. The observational inputs are the same as in \citet{Goradzhanov2025}: the \textsc{MGE} decomposition of the \textsc{DESI} $z$-band image (seven Gaussians per galaxy), which specifies the stellar density distribution, and the \textsc{MaNGA} stellar kinematic maps, represented by Gauss-Hermite moments ($V$, $\sigma$, $h_3$, $h_4$) in Voronoi bins. The problem is solved in the axisymmetric approximation. In \citet{Goradzhanov2025}, the dark halo was described by a modified \textsc{NFW} profile, with its scale radius $r_\mathrm{scale}$ and peak circular velocity $v_\mathrm{circ}$ fitted simultaneously, while a single mass-to-light ratio $\Upsilon$ was applied to the entire stellar \textsc{MGE} model. Here we use a cored isothermal profile, fix the core radius at the effective radius of the galaxy, $r_\mathrm{core}=R_e$, and fit only the halo normalization (Section~\ref{subsec:data_halo}); we also introduce a radial form $\Upsilon(R)$. We therefore construct the dynamical model anew and describe it in the necessary detail. To ensure consistency between the dynamics and spectral synthesis, we refine two input parameters. First, the radial profile of the stellar factor $M/L_\star$ is taken from the same \textsc{E-MILES} models (with the Kroupa Universal IMF \citep{Kroupa2001MNRAS.322..231K}) as the age and metallicity maps, rather than from an independent fit: for each Voronoi bin, the $M/L$ ratio (here and below, in the $z$ band) is recovered from the \nb\ age and metallicity estimates using the same SSP grid, $\Upsilon=\Upsilon(T_\mathrm{SSP},[\mathrm{M/H}])$, yielding a spatial $M/L$ map. Its radial behavior is approximated by a separate factor $\Upsilon_k$ for each of the seven \textsc{MGE} Gaussians (the component-wise $\Upsilon$ method; \citealt{Forstand}), specifying the radially varying \emph{shape} $\Upsilon(R)$ of the stellar density, while the overall amplitude $\Upsilon_0$ remains free and is optimized in the fit (see below). Second, we adjust the absolute photometric normalization of the \textsc{MGE} to the catalog total $z$-band magnitude of the galaxy adopted in the dynamical model. For PGC~35706, this yields a physical $M/L\approx1.3$ (in solar units), of the same order as its own population value of $\approx1.8$ from the same SSP models, providing an independent consistency check within the photometric systematic uncertainties; the younger LEDA~2220522 has a population $M/L$ roughly three times lower ($\approx0.5$). The recovered masses and dark halo constraints agree with the results of \citet{Goradzhanov2025} within these refinements; the fundamentally new content of this work is the stellar population and orbit decomposition and the estimation of SSP-equivalent ages for the stellar populations of the components.

\subsection{Dark halo parameterization and constraints}
\label{subsec:data_halo}

The gravitational potential comprises a stellar component whose mass-to-light ratio $\Upsilon(R)$ varies with radius according to the population profile (the factors $\Upsilon_k$ for the \textsc{MGE} Gaussians described above), a central black hole of mass $M_\bullet = 10^{7}\,M_\odot$ (fixed; within the field of view actually used, the results are insensitive to its value up to $\sim\!10^{8}\,M_\odot$, as expected from the $M_\bullet$--$\sigma$ relation), and a dark halo represented by an isothermal (logarithmic) profile. Within the \textsc{MaNGA} field of view actually used, the data constrain one mass degree of freedom of the halo: simultaneous fitting of its scale and normalization is degenerate. We therefore \emph{fix} the halo core radius at the galaxy effective radius $R_e$ (from the \textsc{MGE} photometry), leaving only the normalization free; the ratio $\Upsilon$ is optimized by scaling velocities by $\sqrt{\Upsilon}$ \citep{Forstand}. We express the halo normalization independently of the profile through the dark matter contribution to the circular velocity at the effective radius, $V_{c,\mathrm{DM}}(R_e)=\sqrt{G\,M_{\mathrm{DM}}(<\!R_e)/R_e}$, because the data constrain the central dark mass substantially better than the asymptotic behavior of the profile. The model thus reduces to a one-dimensional scan over $V_{c,\mathrm{DM}}(R_e)$: an orbit library containing $2\times10^{4}$ orbits is integrated for each value. Its weights are found in a single joint optimization subject to constraints on the stellar density and all four observed kinematic maps: $V$, $\sigma$, $h_3$, and $h_4$. In other words, the kinematic term in the objective function is the sum of the normalized squared residuals of all four moments in all Voronoi bins; separate independent fits to the maps are not performed. Regularization of order unity is applied in solving the problem.

\begin{figure*}[t!]
\includegraphics[width=\linewidth]{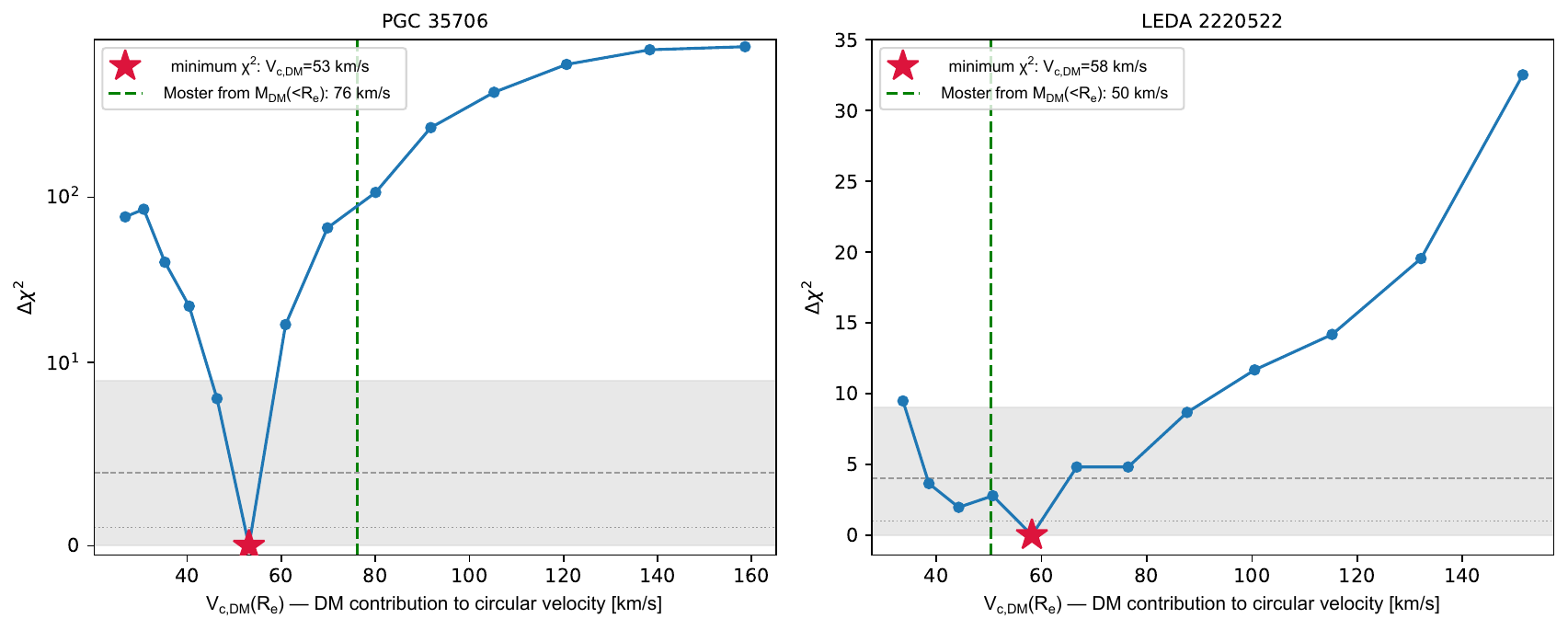}
\caption{$\Delta\chi^2=\chi^2-\chi^2_{\min}$ as a function of the dark matter contribution to the circular velocity at the effective radius, $V_{c,\mathrm{DM}}(R_e)$ (isothermal halo, $r_{\mathrm{core}}=R_e$, optimized $M/L_\star$ factor), for PGC~35706 (left) and LEDA~2220522 (right). The gray band marks $\Delta\chi^2<9$. For PGC~35706, the $\Delta\chi^2$ axis uses a symlog scale (linear for $\Delta\chi^2<10$, logarithmic above) because of the large dynamic range: the minimum at $V_{c,\mathrm{DM}}\approx53$~km/s, with a rise on both sides, confirms that it is an \emph{interior} minimum rather than the edge of the scan. The green dashed line shows the prediction from the $M_\star$--$M_\mathrm{h}$ relation \citep{Moster2013}, positioned according to $M_{\mathrm{DM}}(<\!R_e)$ (by mass, not by $\chi^2$: the \textsc{NFW} and isothermal halo profiles are not directly comparable). For PGC~35706, the minimum is shifted toward a low-mass halo, so the Moster prediction overestimates $M_{\mathrm{DM}}(<\!R_e)$ by about a factor of two; for LEDA~2220522, the bottom of the $\chi^2$ curve is flat.}
\label{fig:chi2vh}
\end{figure*}

\begin{figure*}[t!]
\includegraphics[width=\linewidth]{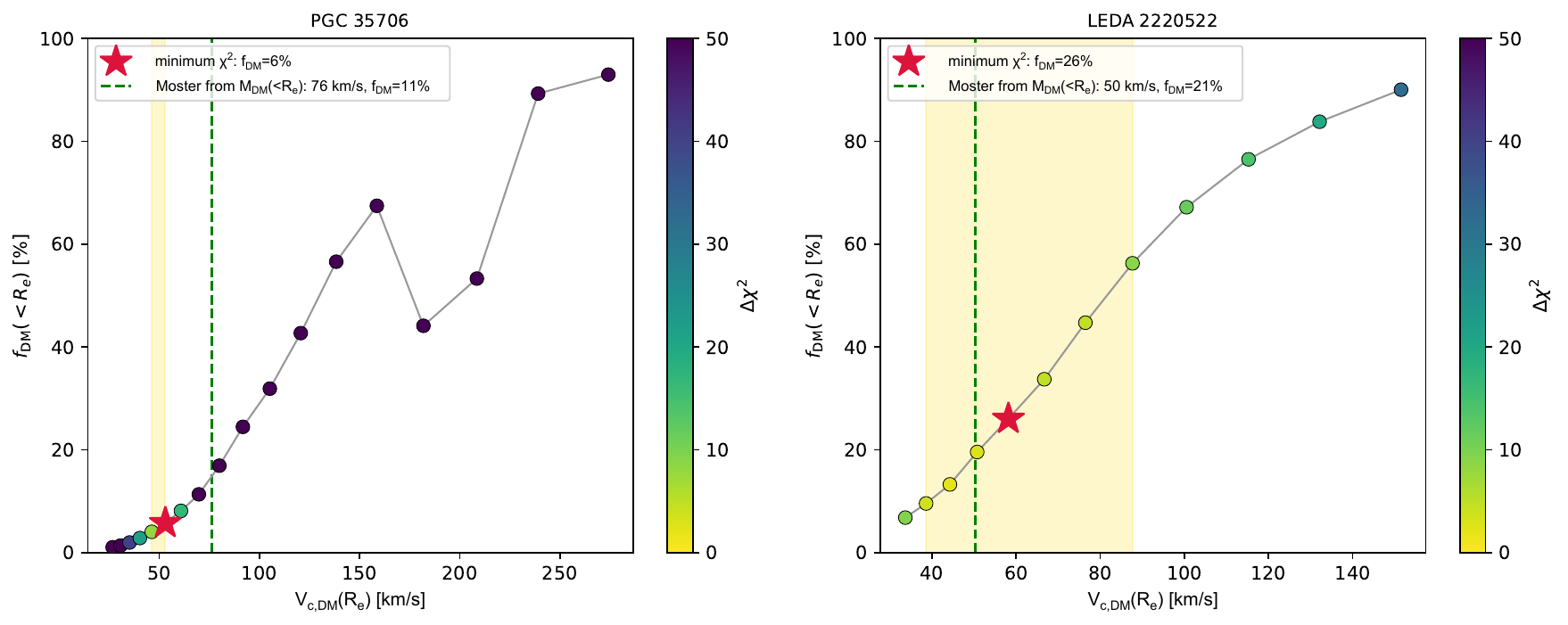}
\caption{Dark mass fraction $f_{\mathrm{DM}}(<\!R_e)$ as a function of $V_{c,\mathrm{DM}}(R_e)$; point colors encode $\Delta\chi^2$. The gold band marks the statistically allowed normalization ($\Delta\chi^2<9$); the green dashed line shows the $M_\star$--$M_\mathrm{h}$ relation \citep{Moster2013} expressed through $M_{\mathrm{DM}}(<\!R_e)$. For LEDA~2220522, the allowed range is $f_{\mathrm{DM}}(<\!R_e)=10$--$56\%$ (the Moster prediction lies within the band); in the adopted model, the $\chi^2$ minimum for PGC~35706 corresponds to $f_{\mathrm{DM}}\approx6\%$.}
\label{fig:fdmvh}
\end{figure*}

The resulting $\chi^2(V_{c,\mathrm{DM}})$ curves (Fig.~\ref{fig:chi2vh}) differ qualitatively between the two galaxies. For LEDA~2220522, the bottom of the $\chi^2$ curve is flat: models with $V_{c,\mathrm{DM}}(R_e)\approx35$--$90$~km/s are practically indistinguishable, and the halo normalization is not determined by the data (the formal minimum occurs at $V_{c,\mathrm{DM}}\approx58$~km/s, $f_{\mathrm{DM}}(<\!R_e)\approx26\%$, but is not statistically distinct). For PGC~35706, in contrast, the central dark mass is much better constrained: $\chi^2$ has a pronounced interior minimum at $V_{c,\mathrm{DM}}(R_e)\approx53$~km/s, favoring a low-mass central halo. The corresponding dark mass fraction within $R_e$ (Fig.~\ref{fig:fdmvh}) is $f_{\mathrm{DM}}(<\!R_e)\approx10$--$56\%$ for LEDA~2220522 (observationally undetermined), but only $f_{\mathrm{DM}}(<\!R_e)\approx6\%$ for PGC~35706. Such a low central dark matter fraction is consistent with values typical of early-type galaxies: according to the ATLAS$^\mathrm{3D}$ survey, at $\log M_\star\approx10.4$--$10.7$ the dark mass fraction within $R_e$ generally does not exceed $\sim\!10\%$ \citep{Cappellari_2013}. Since the stellar population and orbit model takes the orbit distribution as input, rather than the halo normalization itself, this degeneracy has little effect on the subsequent analysis: the orbital decomposition is insensitive to the choice of halo profile---models with isothermal and \textsc{NFW} halos \citep{NFW} yield consistent results---while $f_{\mathrm{DM}}$ and the decomposition itself depend only on mass ratios and are invariant under changes in the absolute luminosity normalization.

We directly tested whether the ionized-gas kinematics could provide an independent constraint using the \Ha\ line-of-sight velocity fields from \textsc{MaNGA DAP}. After selecting spaxels with $S/N(\Ha)\geq5$, we fitted a simple axisymmetric rotating-disk model and constructed folded curves for the two sides of each galaxy. Such an interpretation is not applicable to PGC~35706: the RMS residual of the field is $41$~km/s, compared to a median formal error of $6$~km/s, and the two sides of the curve differ by an average of $25$~km/s, indicating substantial noncircular motions. The field of LEDA~2220522 is much more regular (RMS $4.7$~km/s; difference between the sides $7.7$~km/s), and at the adopted inclination of $i=27^\circ$ the folded curve reaches $\simeq75$--$81$~km/s at radii of $2.5$--$3.5\arcsec$. However, because the inclination is low, a change in $i$ of only $\pm5^\circ$ changes the characteristic velocity estimate from approximately $73$ to $101$~km/s, with almost no change in the fit quality. Moreover, the gas curve constrains the total circular velocity but does not by itself separate the stellar and dark halo contributions. We therefore use it only as a qualitative independent check; incorporating the gas into a quantitative halo constraint would require jointly fitting the inclination, stellar $M/L$, and possible noncircular motions.

We use the cosmologically expected halo normalization from the stellar mass--halo mass relation (abundance matching) as an independently testable prediction rather than a constraint, and make the comparison independently of the profile, using the dark mass within $R_e$ and its contribution to the circular velocity $V_{c,\mathrm{DM}}(R_e)$ rather than the asymptotic maximum circular velocity $V_\mathrm{max}$. This distinction matters: at the same $V_\mathrm{max}$, the \textsc{NFW} and isothermal halo profiles correspond to different central masses, because the \textsc{NFW} rotation curve passes through a maximum at $\sim\!2.2\,r_s$ and declines toward the center, whereas $V_c$ in an isothermal halo increases monotonically toward its asymptotic value. For the massive PGC~35706 ($\log(M_\star/M_\odot)=10.713$ after converting the \textsc{DynPop} catalog mass using $h=0.73$), the relation of \citet{Moster2013}, with the concentration from \citet{DuttonMaccio2014}, gives $M_{\mathrm{DM}}(<\!R_e)\approx2.7\times10^{9}~M_\odot$, or $V_{c,\mathrm{DM}}(R_e)\approx76$~km/s, whereas in our adopted model the $\chi^2$ minimum for PGC~35706 corresponds to $M_{\mathrm{DM}}(<\!R_e)\approx1.3\times10^{9}~M_\odot$ ($V_{c,\mathrm{DM}}(R_e)\approx53$~km/s), a mass approximately half as large. This discrepancy, however, is expected rather than anomalous: low central $f_{\mathrm{DM}}(<\!R_e)$ values are typical of early-type galaxies (cf.\ ATLAS$^\mathrm{3D}$; \citealt{Cappellari_2013}), while cosmologically normalized halos tend to overestimate the \emph{central} dark mass specifically. We treat the magnitude of the discrepancy cautiously: it is comparable to the systematic uncertainty in the central dynamical mass (the choice of amplitude $\Upsilon_0$ and the axisymmetric approximation), which could also change its sign, so we draw no general conclusions about halo normalization in massive galaxies from a single object. For the less massive LEDA~2220522, the prediction ($V_{c,\mathrm{DM}}\approx50$~km/s) falls within the degenerate range and is not constrained by the data.

\subsection{Orbit library and dynamical components}
\label{subsec:data_orbits}

\begin{figure*}[ht!]
\begin{minipage}[h]{0.49\linewidth}
\center{\includegraphics[trim=0cm 0cm 0cm 0cm, clip, width=1\linewidth]{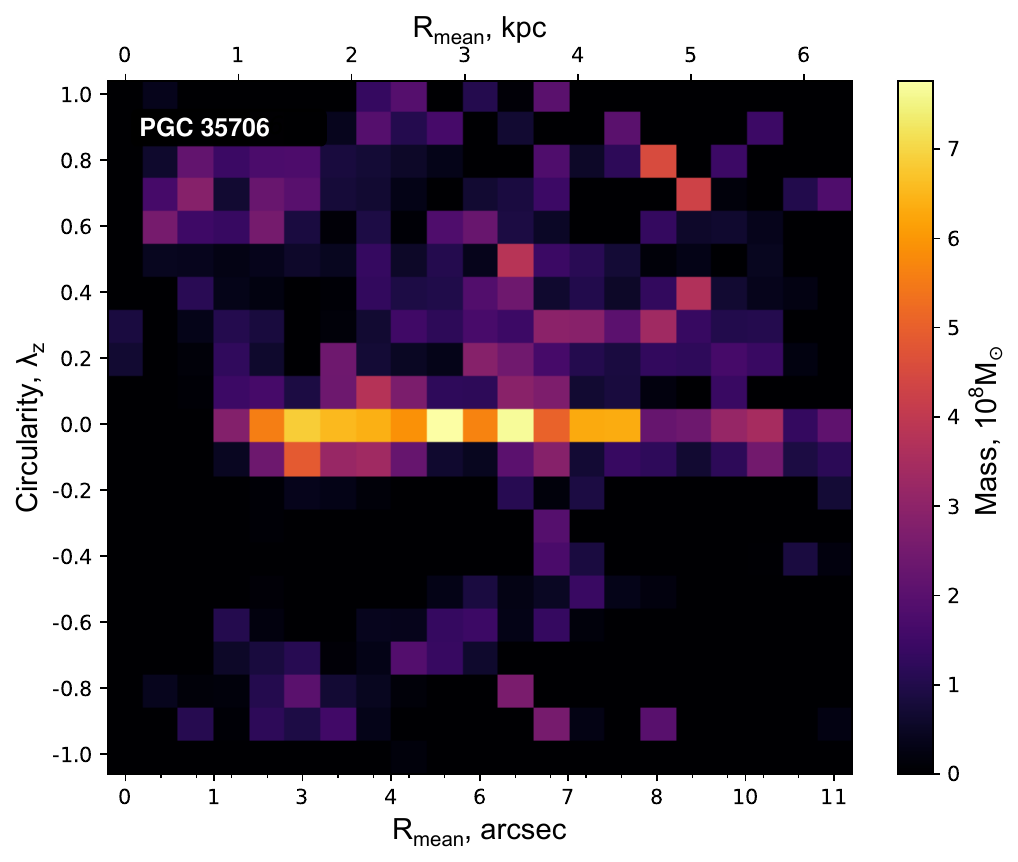}}
\end{minipage}
\hfill
\begin{minipage}[h]{0.49\linewidth}
\center{\includegraphics[trim=0cm 0cm 0cm 0cm, clip, width=1\linewidth]{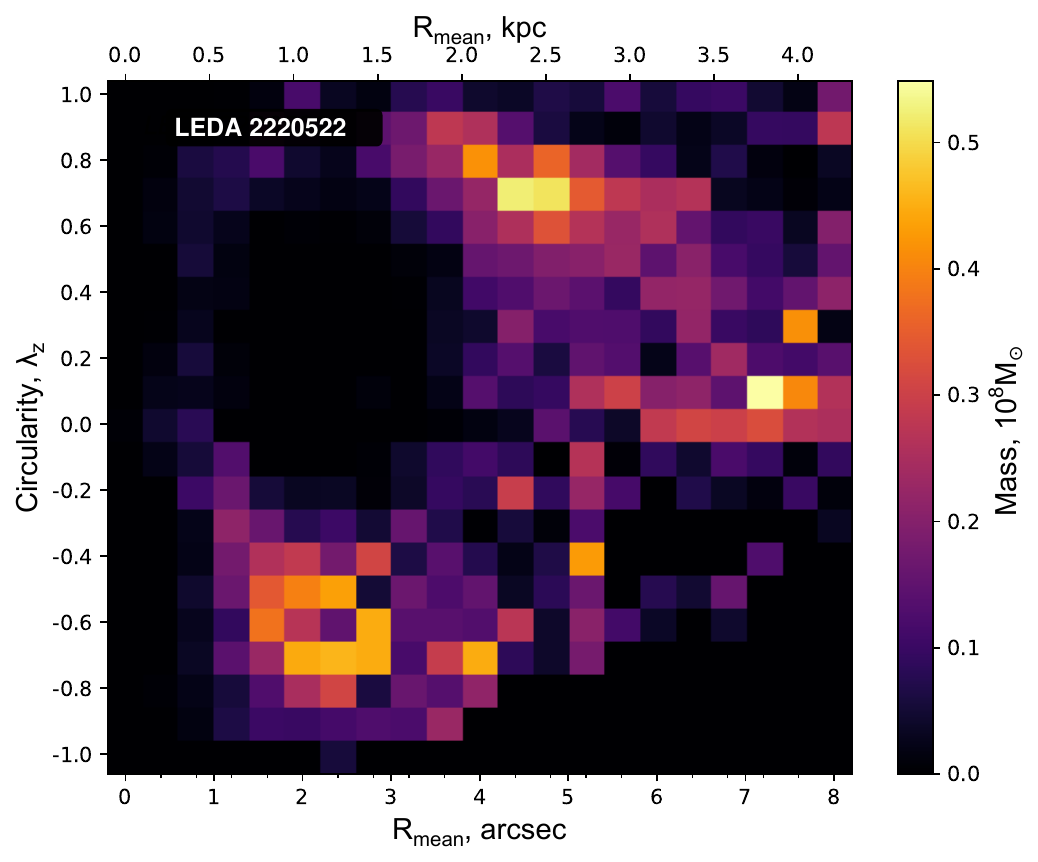}}
\end{minipage}
\caption{Orbital mass distribution in units of
$10^8~M_\odot$ in the $(R_\mathrm{mean},\,\lambda_z)$ plane for PGC~35706 (left) and LEDA~2220522 (right), obtained from the dynamical modeling in this work. The vertical axis shows the orbital circularity parameter $\lambda_z$: values of $|\lambda_z|\lesssim0.35$ correspond to the spheroidal component, $\lambda_z\gtrsim+0.35$ to the co-rotating disk, and $\lambda_z\lesssim-0.35$ to the counter-rotating disk. This distribution is used here as the ``framework'' onto which the stellar population and orbit model (Section~\ref{sec:method}) maps the age and metallicity values.}
\label{ris:orbit_maps}
\end{figure*}

The final orbital weights allow the full orbital mass distribution to be reconstructed in the $(R_\mathrm{mean}, \lambda_z)$ plane, where $\lambda_z \equiv \overline{L_z}/L_\mathrm{circ}(E)$ is the circularity parameter. The two-dimensional orbital mass distributions constructed for PGC~35706 and LEDA~2220522 in this plane are shown in Fig.~\ref{ris:orbit_maps}; they provide the starting point for stellar population and orbit modeling. From this distribution, we identify three kinematically distinct regions, following the scheme of \citet{Goradzhanov2025}. To ensure consistency with the uniform $21\times 21$ grid in the $(R,\lambda_z)$ plane (Section~\ref{sec:method}), the boundaries are chosen so that each third of the $\lambda_z\in[-1,+1]$ axis contains exactly seven cells: the counter-rotating disk ($-1 \le \lambda_z < -0.35$), the spheroidal component ($-0.35 \le \lambda_z < +0.35$), and the co-rotating disk ($+0.35 \le \lambda_z \le +1$). This partition is symmetric about $\lambda_z=0$ and is equivalent in meaning to the scheme adopted by \citet{Goradzhanov2025}; the small differences in the recovered mass fractions from those reported there (at the level of a few percent) arise from the updated dynamical model and the finite orbit sample and are discussed below (Section~\ref{subsec:res_mass}).

In this work, this orbit library serves as the ``framework'' for assigning age and metallicity to the dynamical components: each orbital cell in the $(R,\lambda_z)$ plane receives its own $T_\mathrm{SSP}$ and $[\mathrm{M/H}]_\mathrm{SSP}$ values, extracted from the maps in Fig.~\ref{ris:obs_popmaps} by solving a linear inverse problem (Section~\ref{sec:method}). The corresponding orbital mass distribution in the $(R_\mathrm{mean}, \lambda_z)$ plane for both galaxies is shown in Fig.~\ref{ris:orbit_maps}. The projections of the orbit distribution onto the plane of the sky, which define the spatial ``footprints'' of each cell in the Voronoi binning, are calculated with \textsc{AGAMA} consistently with the stellar kinematic fit.

\begin{figure*}[t!]
\center{\includegraphics[trim=0cm 0cm 0cm 0cm, clip, width=\linewidth]{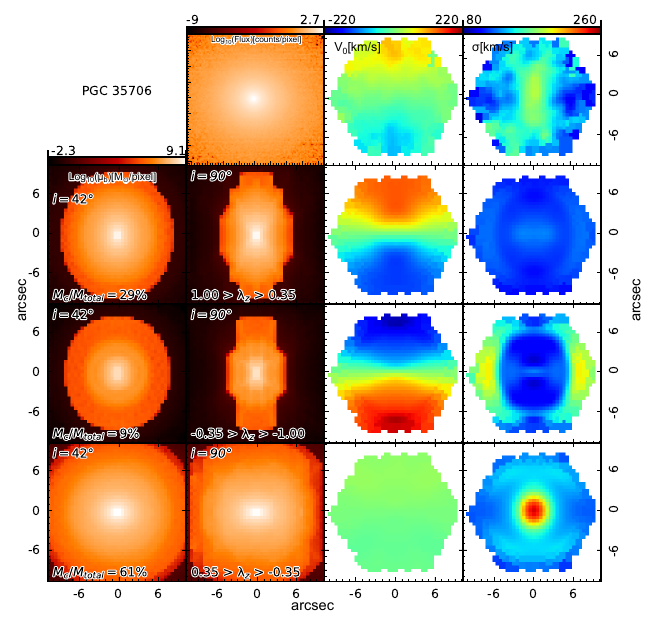}}
\caption{Decomposition of PGC~35706 into three kinematically
distinct components using the dynamical model in this work.
Top row: observed maps. The next three rows show the contributions of the
co-rotating disk ($1.0>\lambda_z>0.35$),
counter-rotating disk ($-0.35>\lambda_z>-1.0$), and
spheroidal component ($0.35>\lambda_z>-0.35$) to the model
image, line-of-sight velocity field, and velocity dispersion field.
The two left columns show the surface density at two disk
inclinations ($i=42^\circ$, as seen from Earth, and $i=90^\circ$, edge-on); the third
column shows the line-of-sight velocity $V_0$, in km/s; the fourth shows the velocity
dispersion $\sigma_0$, in km/s. The contribution of each component to the
total luminosity is indicated in the left panels.}
\label{ris:losvd_PGC}
\end{figure*}

\begin{figure*}[t!]
\center{\includegraphics[trim=0cm 0cm 0cm 0cm, clip, width=\linewidth]{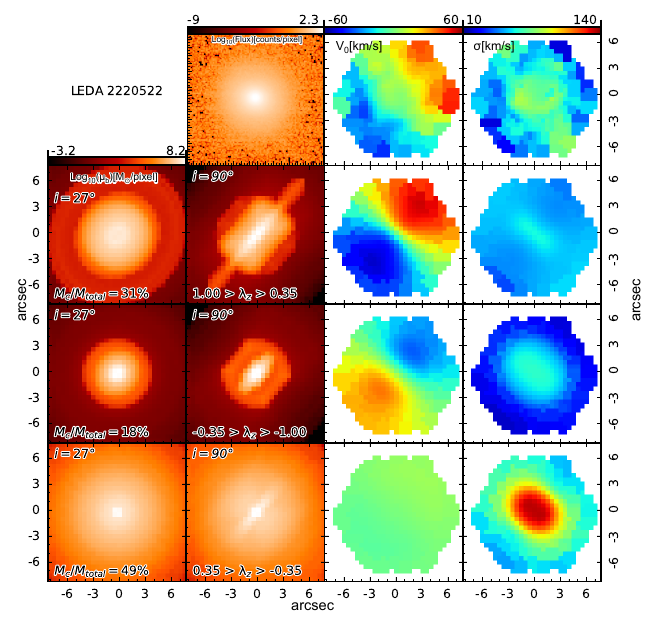}}
\caption{Same as Fig.~\ref{ris:losvd_PGC}, but for
LEDA~2220522; surface density is shown at $i=27^\circ$ (as seen from Earth) and $i=90^\circ$ (edge-on).}
\label{ris:losvd_LEDA}
\end{figure*}

The orbital mass distribution in the $(R_\mathrm{mean},\,\lambda_z)$ plane is accompanied by a spatial and kinematic decomposition of each galaxy into the three identified components. The corresponding component ``footprints'' on the plane of the sky---their surface density, line-of-sight velocity, and velocity dispersion---are shown in Fig.~\ref{ris:losvd_PGC} for PGC~35706 and Fig.~\ref{ris:losvd_LEDA} for LEDA~2220522. They exhibit several characteristic observational signatures of a three-component structure: a compact spheroid with enhanced central velocity dispersion, a two-peaked line-of-sight velocity field and ``$2\sigma$ peaks'' in the counter-rotating disk of PGC~35706, and double counter-rotation visible in the $\sigma$ map of LEDA~2220522. These kinematically distinguishable subsystems generally host stellar populations with different ages and metallicities, which are the subject of Sections~\ref{sec:method} and \ref{sec:results}.

\par\addvspace{0.5\baselineskip}
\subsection{Galaxy environments}
\label{subsec:data_environment}

Characterizing the large-scale environments of the two galaxies is useful for interpreting their formation scenarios. According to the \textsc{SDSS} galaxy group catalog \citep{Tempel2017}, constructed using the friends-of-friends (FoF) method, both galaxies are group members and rank second in luminosity in their respective groups; thus, each group contains a brighter central galaxy. PGC~35706 belongs to a richer group (11 members) than LEDA~2220522 (6 members); both are small groups with dynamical masses of $\sim\!10^{13}$~\Ms\ (rather than clusters; the nearest cluster is $2$--$4$~Mpc away in each case). The group properties are given in Table~\ref{tab:environment}. LEDA~2220522 is offset in line-of-sight velocity from its group by $\approx2\sigma_v$ and has a close satellite at a projected separation of $\approx48$~kpc, consistent with ongoing accretion; for PGC~35706, the most luminous galaxy lies at a projected separation of $\approx343$~kpc, while the nearest faint satellite is at $\approx78$~kpc.

An independent characterization of the large-scale environment is provided by the \textsc{SDSS DR8} filament catalog constructed using the \textsc{Bisous} method \citep{Tempel2014filaments}. In this catalog, PGC~35706 and LEDA~2220522 are assigned to different filaments, with lengths of $11.07$ and $11.73~h^{-1}$~Mpc, respectively; their distances to the nearest filament spine are $0.146$ and $0.071~h^{-1}$~Mpc. Both values are smaller than the filament radius of $0.5~h^{-1}$~Mpc adopted in the catalog, making the catalog filament membership of both systems reasonably secure. However, this geometric association alone does not establish the source or direction of gas accretion on galactic scales; a connection between counter-rotation and material supplied along a filament remains a possible scenario that is not tested in this work.

\begin{table}[ht!]
\begin{tabular}{lcc}
\hline
\hline
 & PGC~35706 & LEDA~2220522 \\
\hline
$N_\mathrm{gal}$            & 11   & 6    \\
$\sigma_v$, km/s            & 246  & 241  \\
$M_{200}$, $10^{13}$~\Ms    & 1.6  & 2.4  \\
$R_{200}$, Mpc              & 0.53 & 0.60 \\
luminosity rank             & 2    & 2    \\
\hline
\end{tabular}
\caption{Properties of the host galaxy groups from the catalog of \citet{Tempel2017}: number of members $N_\mathrm{gal}$, line-of-sight velocity dispersion $\sigma_v$, mass $M_{200}$ and radius $R_{200}$ (model estimates by \citealt{Tempel2017} based on group luminosity and membership rather than $\sigma_v$ alone; approximate for small $N_\mathrm{gal}$), and the luminosity rank of the galaxy under study within its group.}
\label{tab:environment}
\end{table}

\section{METHOD}
\label{sec:method}

The methodology is based on the linear relation between spatially resolved age and metallicity maps and orbital weights in a dynamical model, first formulated by \citet{Poci2019} and further validated using mock data by \citet{Zhu2020}.

\subsection{Formulation of the inverse problem}
\label{subsec:inverse}

Let $i=1,\ldots,N_\mathrm{bin}$ denote the index of a Voronoi bin on the plane of the sky, and $j=1,\ldots,N_\mathrm{cell}$ the index of a cell in the grid of the orbital plane $(R,\lambda_z)$. The dynamical library of each galaxy contains $N_\mathrm{orb}=2\times10^{4}$ individual orbits. For the stellar population and orbit decomposition, these orbits are grouped by their mean radius and circularity into a uniform grid of $N_R \times N_{\lambda_z} = 21 \times 21$ in the dimensionless coordinates $R/R_\mathrm{max}$ and $\lambda_z\in[-1,1]$; thus, $N_\mathrm{cell} = 441$. The dynamical model (Section~\ref{subsec:data_orbits}) defines the projection matrix
\begin{equation}
w_{ij} = \sum_{k \in \mathrm{cell}(j)} \omega_k\,\Pi_{ik},
\label{eq:weights}
\end{equation}
where $\omega_k$ is the set of individual orbital weights giving the best $\chi^2$, which determines their luminosity contributions, $\Pi_{ik}$ is the projection of the $k$th orbit onto the $i$th bin on the plane of the sky, and the sum runs over all orbits $k$ falling within the $(R,\lambda_z)$ cell with index $j$. By construction, $w_{ij}$ represents the fraction of the light flux in the $i$th bin contributed by the orbits of cell $j$.

The observed age in the $i$th bin, $T^\mathrm{obs}_i \equiv T_\mathrm{SSP}(i)$, can be represented as a light-weighted sum:
\begin{equation}
T^\mathrm{obs}_i \;=\;
\frac{\sum_j w_{ij}\,\phi^{T}_j}{\sum_j w_{ij}},
\label{eq:age_model}
\end{equation}
where $\phi^{T}_j$ is the unknown mean stellar population age in cell $j$. An analogous relation is written for the metallicity $[\mathrm{M/H}]^\mathrm{obs}_i$ (with unknowns $\phi^{Z}_j$) and for the mass-to-light ratio: the observed $(M/L)^\mathrm{obs}_i=\Upsilon(T^\mathrm{obs}_i,[\mathrm{M/H}]^\mathrm{obs}_i)$, calculated using the same \textsc{E-MILES} grid, defines a third system of the same form with unknown cell values $\phi^{\Upsilon}_j$. As in \citet{Poci2019}, the solutions for $\boldsymbol\phi^T$ and $\boldsymbol\phi^Z$ are found independently; the covariance between SSP parameters arising during \nb\ fitting is not explicitly included in our procedure. After normalizing the rows of the weight matrix by $\sum_j w_{ij}$, relation~\eqref{eq:age_model} takes the form of a linear system:
\begin{equation}
\begin{aligned}
\mathbf{W}\,\boldsymbol\phi^T
    &= \mathbf{T}^\mathrm{obs},\\
\mathbf{W}\,\boldsymbol\phi^Z
    &= \mathbf{Z}^\mathrm{obs},\\
\mathbf{W}\,\boldsymbol\phi^\Upsilon
    &= \boldsymbol\Upsilon^\mathrm{obs}.
\end{aligned}
\label{eq:linear_system}
\end{equation}
The third equality in the system refers to the mass-to-light ratio map. Below, $\boldsymbol\phi^\gamma$, $\gamma\in\{T,Z,\Upsilon\}$, is used as a general notation for any of the three independently reconstructed fields: age, metallicity, and $M/L$.
Here $\mathbf{W}$ is an $N_\mathrm{bin}\times N_\mathrm{cell}$ matrix; $\boldsymbol\phi^T, \boldsymbol\phi^Z\in\mathbb{R}^{N_\mathrm{cell}}$ are vectors of unknown cell ages and metallicities; and $\mathbf{T}^\mathrm{obs}, \mathbf{Z}^\mathrm{obs} \in \mathbb{R}^{N_\mathrm{bin}}$ are the observed maps. The system dimensions differ between the galaxies: there are $N_\mathrm{bin}=676$ Voronoi bins for PGC~35706 (the 37-fiber bundle) and $N_\mathrm{bin}=263$ for LEDA~2220522 (19 fibers), while the numbers of orbit-occupied $(R,\lambda_z)$ cells (out of $441$ grid cells) are $306$ and $354$, respectively. Thus, PGC~35706 has substantially more equations than unknowns, whereas LEDA~2220522 has slightly more unknowns than equations. Moreover, the effective conditioning of $\mathbf{W}$ is poorer than the nominal ratio of its dimensions suggests: without regularization, $77$--$93\%$ of the cells reach the bounds, depending on the galaxy and the quantity being reconstructed (Section~\ref{subsec:bvls}), indicating a deficiency in the effective rank of the matrix (many cells are weakly constrained by the data). The solution is therefore understood in the regularized least-squares sense, and smoothness regularization is important for the more sparsely covered LEDA~2220522: the narrowness of $\sigma_\mathrm{MC}$ partly reflects the imposed smoothness, rather than the information content of the data alone (see Sections~\ref{subsec:bvls}, \ref{subsec:res_mc}).

\subsection{Bounded linear problem}
\label{subsec:bvls}

The age and metallicity values in each cell are subject to lower and upper bounds:
\begin{equation}
T^\mathrm{min} \le \phi^{T}_j \le T^\mathrm{max},
\qquad
Z^\mathrm{min} \le \phi^{Z}_j \le Z^\mathrm{max},
\label{eq:bounds}
\end{equation}
where the bounds are defined not by the edges of the \textsc{E-MILES} grid, but by the range of observed values in the input maps, with a small margin: for PGC~35706, $[T^\mathrm{min},T^\mathrm{max}]=[1.8,\,14.1]$~Gyr and $[Z^\mathrm{min},Z^\mathrm{max}]=[-0.6,\,+0.4]$~dex; for LEDA~2220522, $[1.0,\,5.5]$~Gyr and $[-1.0,\,+0.3]$~dex. Tying the bounds to the data range rather than the library edges is important: for LEDA~2220522, the \textsc{E-MILES} metallicity bounds span almost twice the observed range, and using them literally drives some weakly constrained cells to unphysically low $[\mathrm{M/H}]$ in the bounded fit. Subject to~\eqref{eq:bounds}, system~\eqref{eq:linear_system} is solved as a bounded-variable least-squares (\textsc{BVLS}; \citealt{StarkParker1995}) problem using the \texttt{scipy.optimize.lsq\_linear} function with \texttt{method=\textquotesingle bvls\textquotesingle} (implementing the Stark--Parker algorithm) from the \textsc{SciPy} library \citep{Virtanen2020}; smoothness regularization (see below) is introduced by appending rows $\lambda L$ to the system, rather than being implemented within the solver. The solutions for $\boldsymbol\phi^T$, $\boldsymbol\phi^Z$, and $\boldsymbol\phi^{\Upsilon}$ are found independently; for the last of these, the bounds $[\Upsilon^\mathrm{min},\Upsilon^\mathrm{max}]$ are set by the range of the $M/L$ map (with the same small margin).

To stabilize the solution in weakly constrained regions, where \textsc{BVLS} would otherwise drive values to the bounds, smoothness regularization in the $(R,\lambda_z)$ plane is added to system~\eqref{eq:linear_system}. The matrix $L$ is a discrete graph Laplacian of the occupied region of the phase-space grid: $L_{jj}=n_j$, where $n_j$ is the number of actual neighbors of cell $j$ along $R$ and $\lambda_z$, and $L_{jk}=-1$ for each neighboring pair. Thus, the elements of each row sum to zero, a constant field is not penalized, and no implicit cells with zero values are introduced at the irregular boundary of the region. Rows $\lambda L$ with weight $\lambda=0.01$ are appended to the system.

To compare the quality of solutions for different $\lambda$, we use the mean error-weighted squared residual over the $N_\mathrm{valid}$ valid observational bins only,
\begin{equation}
\chi^2_\mathrm{red}=\frac{1}{N_\mathrm{valid}}
\sum_{i\in\mathrm{valid}}
\left(\frac{(\mathbf W\boldsymbol\phi)_i-y_i^\mathrm{obs}}{\sigma_i}\right)^2,
\label{eq:chi2red_population}
\end{equation}
excluding the regularization rows. Here the subscript ``red'' denotes normalization by the number of observations used, rather than formal division by the number of degrees of freedom $N_\mathrm{valid}-N_\mathrm{par}$: the effective number of parameters in a bounded regularized solution is not uniquely defined. The value of $\lambda$ was selected by rescanning for the ``knee'' in $\chi^2_\mathrm{red}(\lambda)$ after constructing the Laplacian with the same cell ordering as the columns of $\mathbf{W}$. The limiting case is the PGC~35706 age map: $\chi^2_\mathrm{red}$ increases from $58.82$ at $\lambda=0$ to $59.83$ at $\lambda=0.01$, but already reaches $62.32$ at $\lambda=0.03$. At $\lambda=0.01$, the fraction of cells at the bounds of the allowed range decreases to $39\%$ and $30\%$ for age and metallicity in PGC~35706, and to $37\%$ and $38\%$, respectively, for LEDA~2220522. The same value, $\lambda=0.01$, is adopted for both galaxies. This choice based on the knee is heuristic (rather than satisfying a formal optimality criterion such as GCV); the contribution of the imposed smoothness to the narrowness of $\sigma_\mathrm{MC}$ is discussed separately in Sections~\ref{subsec:res_mc} and \ref{subsec:disc_limits}.

It should be emphasized that, within each individual $(R,\lambda_z)$ cell, the solution assumes a single mean age and a single mean metallicity. In other words, the model constructs a mosaic field on the plane of the sky with values $\phi^T_j, \phi^Z_j$ shared by all bins associated with a given orbital cell. This constraint is fully analogous to the approach of \citet{Poci2019} and imposes a fundamental limit on the resolving power of the method: unresolved mixtures of multiple populations or sharp chemical gradients within a single cell cannot be reproduced and appear as systematic residuals between the model and observed maps (Section~\ref{subsec:residuals}). The choice of a $21\times 21$ grid represents a compromise between resolving small kinematically distinct substructures in the counter-rotating disk and maintaining acceptable conditioning of the system, given the actual number of Voronoi bins and the typical errors in $T_\mathrm{SSP}$ and $[\mathrm{M/H}]_\mathrm{SSP}$.

\subsection{Component identification in the
\texorpdfstring{$(R,\lambda_z)$}{(R, lambda\_z)} plane}
\label{subsec:components}

The recovered fields $\phi^T_j$ and $\phi^Z_j$ on the $(R,\lambda_z)$ grid naturally combine into kinematically distinct components according to the partition of the $\lambda_z$ axis into three groups with equal numbers of cells (Section~\ref{subsec:data_orbits}): the spheroidal component ($-0.35 \le \lambda_z < +0.35$), the co-rotating disk ($+0.35 \le \lambda_z \le +1$), and the counter-rotating disk ($-1 \le \lambda_z < -0.35$). Since the population parameters are constrained by the data only within the \textsc{MaNGA} field of view, we use the mass of each cell's projection within the observed field for averaging, rather than the total mass of the orbits belonging to the cell. The matrix $W_{ij}=w_{ij}/\sum_k w_{ik}$ used in system~\eqref{eq:linear_system} still describes the normalized luminosity contribution of cell $j$ to observational bin $i$. For mass weighting, let $M_i^\mathrm{proj}$ denote the total projected mass in bin $i$, and $\widetilde m_{ij}=W_{ij}M_i^\mathrm{proj}$ the mass contribution of cell $j$ to this bin. The cell mass within the field of view is then
\begin{equation}
M_j^\mathrm{FoV}=\sum_{i=1}^{N_\mathrm{bin}}\widetilde m_{ij}.
\label{eq:cell_fov_mass}
\end{equation}
The mean age, metallicity, and mass-to-light ratio for component $C\in\{\mathrm{sph}, \mathrm{co}, \mathrm{counter}\}$ are calculated using these projected masses (the expression for $M/L$ is fully analogous, with $\phi^T_j\to\phi^{\Upsilon}_j$):
\begin{equation}
\begin{aligned}
\langle T\rangle_C
    &=\frac{\sum_{j\in C} M_j^\mathrm{FoV}\,\phi^T_j}
            {\sum_{j\in C} M_j^\mathrm{FoV}},\\
\langle Z\rangle_C
    &=\frac{\sum_{j\in C} M_j^\mathrm{FoV}\,\phi^Z_j}
            {\sum_{j\in C} M_j^\mathrm{FoV}}.
\end{aligned}
\label{eq:comp_averages}
\end{equation}
Radial profiles are constructed directly on the plane of the sky. A projected surface density map $\Sigma_j(x,y)$ is reconstructed for each cell of the dynamical model; the field of view is divided into 21 circular annuli $\mathcal A_a$ of equal width in projected radius $R_\mathrm{proj}=\sqrt{x^2+y^2}$. The mass of cell $j$ in annulus $a$ is defined as
\begin{equation}
M_{aj}=\int_{\mathcal A_a\cap\mathrm{FoV}}\Sigma_j(x,y)\,\mathrm{d}A,
\label{eq:annular_cell_mass}
\end{equation}
and the component profile is then calculated as
\begin{equation}
\begin{aligned}
\langle T\rangle_C(R_a)
    &=\frac{\sum_{j\in C} M_{aj}\,\phi^T_j}
            {\sum_{j\in C} M_{aj}},\\
\langle Z\rangle_C(R_a)
    &=\frac{\sum_{j\in C} M_{aj}\,\phi^Z_j}
            {\sum_{j\in C} M_{aj}}.
\end{aligned}
\label{eq:radial_profile}
\end{equation}
The maximum covered radius, that is, the distance from the center to the most distant valid spaxel in the actual hexagonal \textsc{MaNGA} field, is $R_\mathrm{proj}=10.12\arcsec$ for PGC~35706 and $7.83\arcsec$ for LEDA~2220522. The outer circular annuli are therefore filled only where they intersect the observed field; in particular, the outermost annulus of LEDA~2220522 contains only $0.01\%$ of the total projected mass within the field. If the mass of the component under consideration in a given annulus is less than $1\%$ of the total annular mass, the corresponding profile segment is retained for completeness but shown as a dotted line because it is weakly constrained by the projected data.

\subsection{Reconstruction of model maps and assessment of residuals}
\label{subsec:residuals}

Once the values $\phi^T_j, \phi^Z_j$ have been determined by solving~\eqref{eq:linear_system}--\eqref{eq:bounds}, the two-dimensional model age and metallicity maps are reconstructed by direct projection onto the Voronoi bins:
\begin{equation}
T^\mathrm{mod}_i =
\frac{\sum_j w_{ij}\,\phi^T_j}{\sum_j w_{ij}},
\qquad
Z^\mathrm{mod}_i =
\frac{\sum_j w_{ij}\,\phi^Z_j}{\sum_j w_{ij}}.
\label{eq:model_maps}
\end{equation}
The residuals $\Delta T_i = T^\mathrm{mod}_i - T^\mathrm{obs}_i$ and $\Delta Z_i = Z^\mathrm{mod}_i - Z^\mathrm{obs}_i$ characterize the model's ability to reproduce the observed spatial variations in age and metallicity. Below, a spatially coherent residual means a connected region of several adjacent independent bins with deviations of the same sign, rather than an isolated outlier. Such regions may indicate either inadequacy of the assumption of a single age/metallicity within individual orbital cells (Section~\ref{subsec:bvls}) or limitations of the adopted axisymmetric dynamical model. The residual maps are presented and discussed together with the model maps in Section~\ref{sec:results}.

\subsection{Monte Carlo assessment of stability}
\label{subsec:mc}

The finite precision of the inferred values $\phi^T_j$ and $\phi^Z_j$ is limited by the errors in the input maps $T^\mathrm{obs}_i$ and $Z^\mathrm{obs}_i$, which are estimated for each bin by \nb\ from the covariance matrix of the $\chi^2$-minimization parameters. To propagate these errors to the $(R,\lambda_z)$ plane and then to the resulting component means and radial profiles, we use a Monte Carlo procedure similar in spirit to that of \citet{Poci2019}.

For each of the $N_\mathrm{MC} = 50$ realizations, the observed values are perturbed independently in each bin according to their individual errors $\sigma^T_i$ and $\sigma^Z_i$. To ensure positive ages, we use a lognormal distribution with expectation $T_i^\mathrm{obs}$ and standard deviation $\sigma_i^T$:
\begin{equation}
\begin{aligned}
\widetilde{T}^{\mathrm{obs},(s)}_i
    &\sim \mathrm{LogNormal}(\mu_i,s_i^2),\\
s_i^2
    &=\ln\!\left[1+\left(\frac{\sigma_i^T}{T_i^\mathrm{obs}}\right)^2\right],\\
\mu_i
    &=\ln T_i^\mathrm{obs}-\frac{s_i^2}{2}.
\end{aligned}
\label{eq:mc_perturb}
\end{equation}
whereas metallicity is perturbed normally, $\widetilde{Z}^{\mathrm{obs},(s)}_i=Z_i^\mathrm{obs}+\eta_i^{Z,(s)}$, $\eta_i^{Z,(s)}\sim\mathcal N(0,(\sigma_i^Z)^2)$. The age and metallicity perturbations are uncorrelated with one another. For each realization $s=1,\ldots,N_\mathrm{MC}$, the bounded linear problem~\eqref{eq:linear_system}--\eqref{eq:bounds} is solved independently, and radial profiles and component means are calculated using~\eqref{eq:comp_averages}--\eqref{eq:radial_profile}. The $M/L$ map is also recalculated from the perturbed $T_\mathrm{SSP}$ and $[\mathrm{M/H}]_\mathrm{SSP}$ using the \textsc{E-MILES} grid, after which its separate \textsc{BVLS} decomposition is repeated. The resulting distribution of solutions provides empirical estimates of the mean and standard deviation for each reconstructed quantity. The corresponding $\pm1\sigma_\mathrm{MC}$ band is shown in the radial-profile figures (Section~\ref{sec:results}), while maps of the ensemble mean and standard deviation are used to assess the stability of the two-dimensional reconstructions. We specifically emphasize that the resulting $\sigma_\mathrm{MC}$ reflects only the statistical error component associated with noise in the input maps and does not include systematic effects related to the choice of dynamical model, regularization, SSP grid, or the single-population assumption for orbital cells; these effects are discussed in Section~\ref{sec:discussion}.

In all realizations, $\boldsymbol\phi^T$ and $\boldsymbol\phi^Z$ are reconstructed independently. The $M/L$ map is then recalculated from the same perturbed pair of input maps using the \textsc{E-MILES} grid, and the separate \textsc{BVLS} decomposition for $\boldsymbol\phi^\Upsilon$ is repeated. Thus, the MC ensemble and the standard deviations reported in Table~\ref{tab:agemet} apply to all three fields $\boldsymbol\phi^\gamma$, $\gamma\in\{T,Z,\Upsilon\}$.

\section{RESULTS}
\label{sec:results}

This section presents the results of stellar population and orbit modeling for PGC~35706 and LEDA~2220522: the mass fractions of the three kinematic components recovered by integrating the orbital weights, two-dimensional model age and metallicity maps, residuals between the model and observed maps, and component radial profiles $T_\mathrm{SSP}(R)$ and $[\mathrm{M/H}]_\mathrm{SSP}(R)$ with Monte Carlo uncertainty estimates. The physical interpretation of these results is discussed in Section~\ref{sec:discussion}.

\subsection{Kinematic component mass fractions within the
MaNGA field of view}
\label{subsec:res_mass}

Summing the projected masses $M_j^\mathrm{FoV}$ over the three regions in the $(R,\lambda_z)$ plane gives the following mass fractions within the \textsc{MaNGA} field actually observed (Table~\ref{tab:fractions}). In PGC~35706, the spheroidal component accounts for $60.3\%$, the co-rotating disk for $30.9\%$, and the counter-rotating disk for $8.8\%$: the spheroid dominates, and the counter-rotating disk is of low mass. In LEDA~2220522, the contributions are more evenly distributed, but the spheroid is again the largest component: $51.4\%$, $31.8\%$, and $16.8\%$, respectively. These estimates refer specifically to the projection within the \textsc{MaNGA} field, rather than to the total mass of the corresponding orbit families; their proximity to the results of \citet{Goradzhanov2025} shows that restricting the analysis to the observed field does not change the qualitative three-component structure.

\begin{table}
\caption{Component mass fractions within the \textsc{MaNGA} field of view actually used (isothermal halo, Kroupa Universal IMF).}
\label{tab:fractions}
\centering
\resizebox{\columnwidth}{!}{%
\begin{tabular}{lccc}
\hline
galaxy & spheroid & co-rotating disk & counter-rotating disk \\
\hline
PGC~35706    & 60.3\% & 30.9\% & 8.8\% \\
LEDA~2220522 & 51.4\% & 31.8\% & 16.8\% \\
\hline
\end{tabular}}
\end{table}

Both galaxies thus exhibit a three-component structure, but their component mass ratios differ substantially. In PGC~35706, the counter-rotating disk has $0.29$ times the mass of the co-rotating disk and less than one tenth of the total mass within the field of view. In LEDA~2220522, the mass ratio of the counter-rotating to co-rotating disk is $0.53$, and together the two disk components contain $48.6\%$ of the mass, comparable to the contribution of the spheroid ($51.4\%$).

\subsection{Two-dimensional model maps and residuals}
\label{subsec:res_maps}

\begin{figure*}[ht!]
\includegraphics[width=\linewidth]{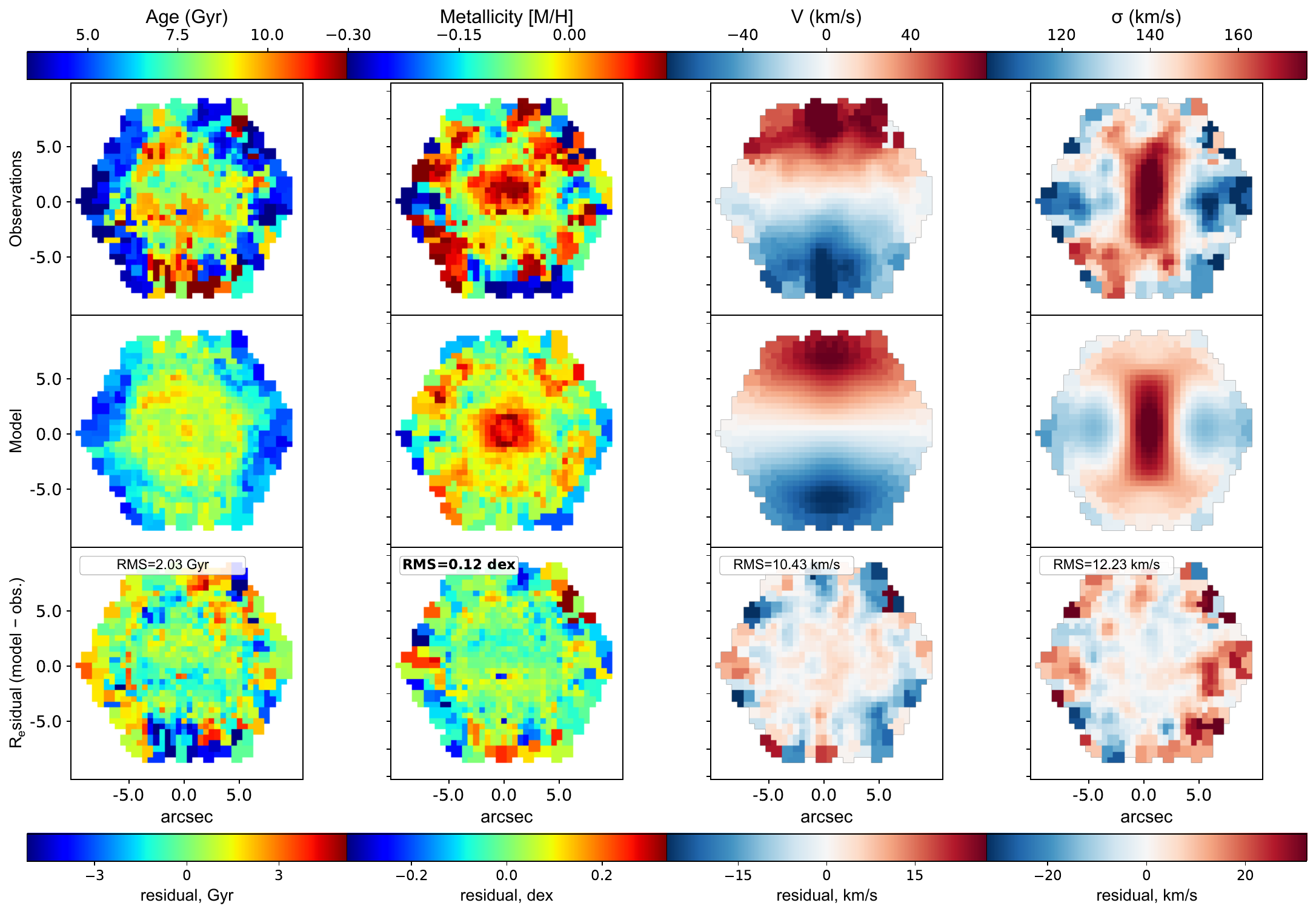}
\caption{Comparison of observed (\textit{top row}) and
model (\textit{middle row}) two-dimensional maps of light-weighted age, metallicity $[\mathrm{M/H}]$, line-of-sight velocity $V$, and line-of-sight velocity dispersion $\sigma$ for PGC~35706. Bottom row: model-minus-observation residuals $X^\mathrm{mod} - X^\mathrm{obs}$. Age is given in Gyr, and $V$ and $\sigma$ in km/s. The axes are in arcseconds from the photometric center. The most prominent local residual in the $[\mathrm{M/H}]$ map occurs in two adjacent bins south of the center and is discussed in the text.}
\label{ris:PGC_residuals}
\end{figure*}

The model age map of PGC~35706 (Fig.~\ref{ris:PGC_residuals}) shows a weak ring-like enhancement at $R_\mathrm{proj}\simeq3$--$4\arcsec$: the mean age in circular annuli here is $8.7$--$8.8$~Gyr, compared to $\simeq8.4$--$8.5$~Gyr at $R_\mathrm{proj}\simeq1.5$--$2.5\arcsec$. The observed map also shows an increase in mean age over the same radial range, but it is distributed among individual bins and does not form such a continuous ring. The more pronounced axial coherence of the model structure is expected from the axisymmetric orbital projections and regularization, which smooth azimuthal fluctuations. We therefore regard it as a smoothed representation of a weak radial feature in the input map, rather than independent evidence for a physically distinct stellar ring.

\begin{figure*}[ht!]
\includegraphics[width=\linewidth]{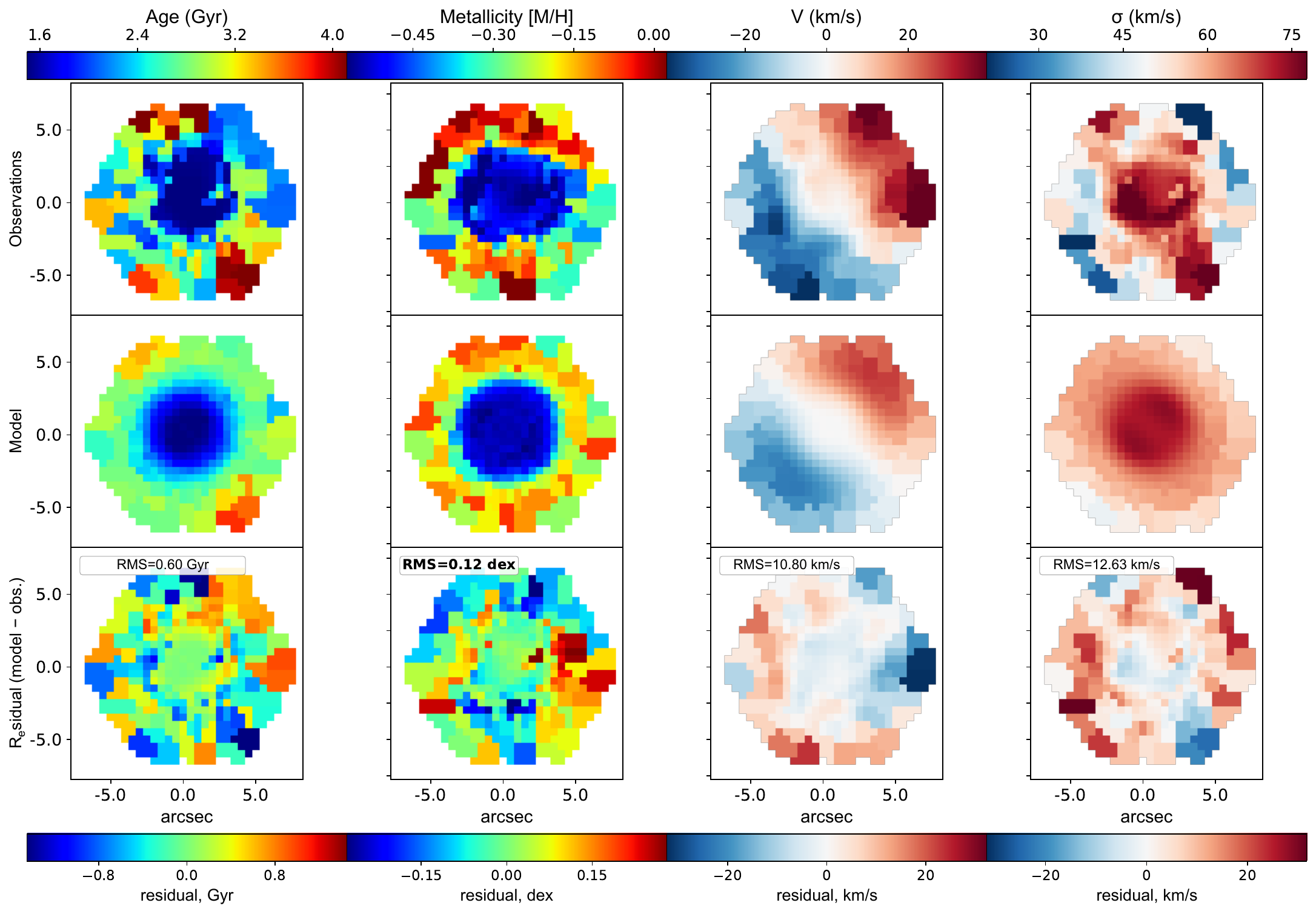}
\caption{Same as Fig.~\ref{ris:PGC_residuals}, but for
LEDA~2220522.}
\label{ris:LEDA_residuals}
\end{figure*}

The model age and metallicity maps reconstructed using~\eqref{eq:model_maps} are shown for both galaxies together with the original observed maps and residual maps in Figs.~\ref{ris:PGC_residuals} (PGC~35706) and \ref{ris:LEDA_residuals} (LEDA~2220522); the same figures show analogous comparisons for the line-of-sight velocity $V$ and velocity dispersion $\sigma$ fields used in fitting the dynamical model (Section~\ref{subsec:data_dyn_input}).

A quantitative check of the central region of PGC~35706 does not confirm large-scale spatial coherence of the $[\mathrm{M/H}]$ residuals (Fig.~\ref{ris:PGC_residuals}). Of the 49 independent bins within $R\leq2\arcsec$, 22 have negative residuals and 27 have positive residuals; the median $\Delta Z$ is $+0.002$~dex, and the RMS residual is $0.068$~dex. The most prominent local discrepancy comprises only two adjacent bins south of the center, where $\Delta Z=-0.22$ and $-0.24$~dex. It may be associated with averaging a steep local $[\mathrm{M/H}]$ gradient within the adopted $(R,\lambda_z)$ grid, or with an unaccounted-for dependence of $[\mathrm{M/H}]$ on age within a single orbital cell; however, the data are insufficient to distinguish these explanations. For LEDA~2220522, the model reproduces the central rise in velocity dispersion, but smooths local asymmetries and variations in $\sigma$ at the edge of the field (Fig.~\ref{ris:LEDA_residuals}); the RMS residual is $12.6$~km/s for $\sigma$ and $10.8$~km/s for $V$. These differences should be taken into account as one of the limitations of the axisymmetric dynamical model.

The model reproduces the structure of the age maps for both galaxies: the spatial residual patterns show no pronounced large-scale trends (bottom row, first column), and the RMS age residual is $\sim\!0.6$~Gyr for LEDA~2220522 and $\sim\!2$~Gyr for the older PGC~35706 (in the latter, it is determined mainly by noise in the observed map at large radii). The situation is more complex for the metallicity maps. In LEDA~2220522, the $[\mathrm{M/H}]$ residuals are spatially random over most of the field of view (Fig.~\ref{ris:LEDA_residuals}), but the model systematically smooths local $[\mathrm{M/H}]$ variations present in the observations: the mean deviation is close to zero, with a small positive offset, and individual pixels reach amplitudes of $\sim\!0.3$~dex. This limitation follows directly from the single-$[\mathrm{M/H}]$ approximation within each orbital cell (Section~\ref{subsec:bvls}).

\subsection{Radial age and metallicity profiles of the
components}
\label{subsec:res_profiles}

\begin{figure*}[ht!]
\includegraphics[width=0.95\linewidth]{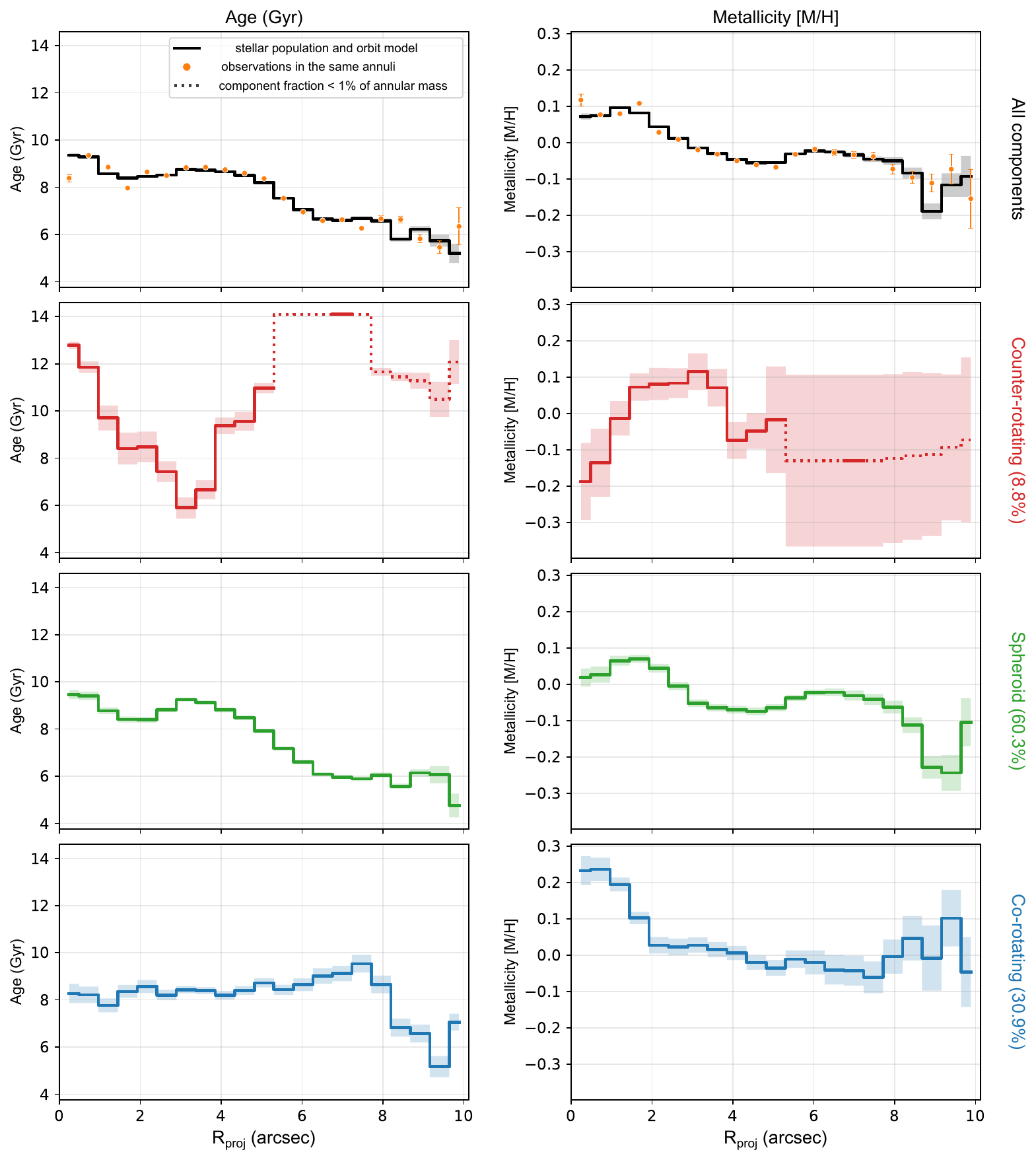}
\caption{Radial age $T_\mathrm{SSP}(R_\mathrm{proj})$ (left column) and metallicity $[\mathrm{M/H}]_\mathrm{SSP}(R_\mathrm{proj})$ (right column) profiles for PGC~35706. Top row: the total profile, weighted by projected mass in each circular annulus within the \textsc{MaNGA} field; orange points with formal errors show the observed SSP-equivalent values averaged in the same annuli and with the same projected-mass weights. Below are the separate profiles of the counter-rotating disk (red, $8.8\%$ of the mass in the field), spheroid (green, $60.3\%$), and co-rotating disk (blue, $30.9\%$). The thick line shows the best-fitting solution; the band is $\pm1\sigma$, estimated from $N_\mathrm{MC}=50$ Monte Carlo realizations. Segments where the component contains less than $1\%$ of the total annular mass are shown as dotted lines and are considered poorly conditioned.}
\label{ris:PGC_radprof}
\end{figure*}

\begin{figure*}[ht!]
\includegraphics[width=0.95\linewidth]{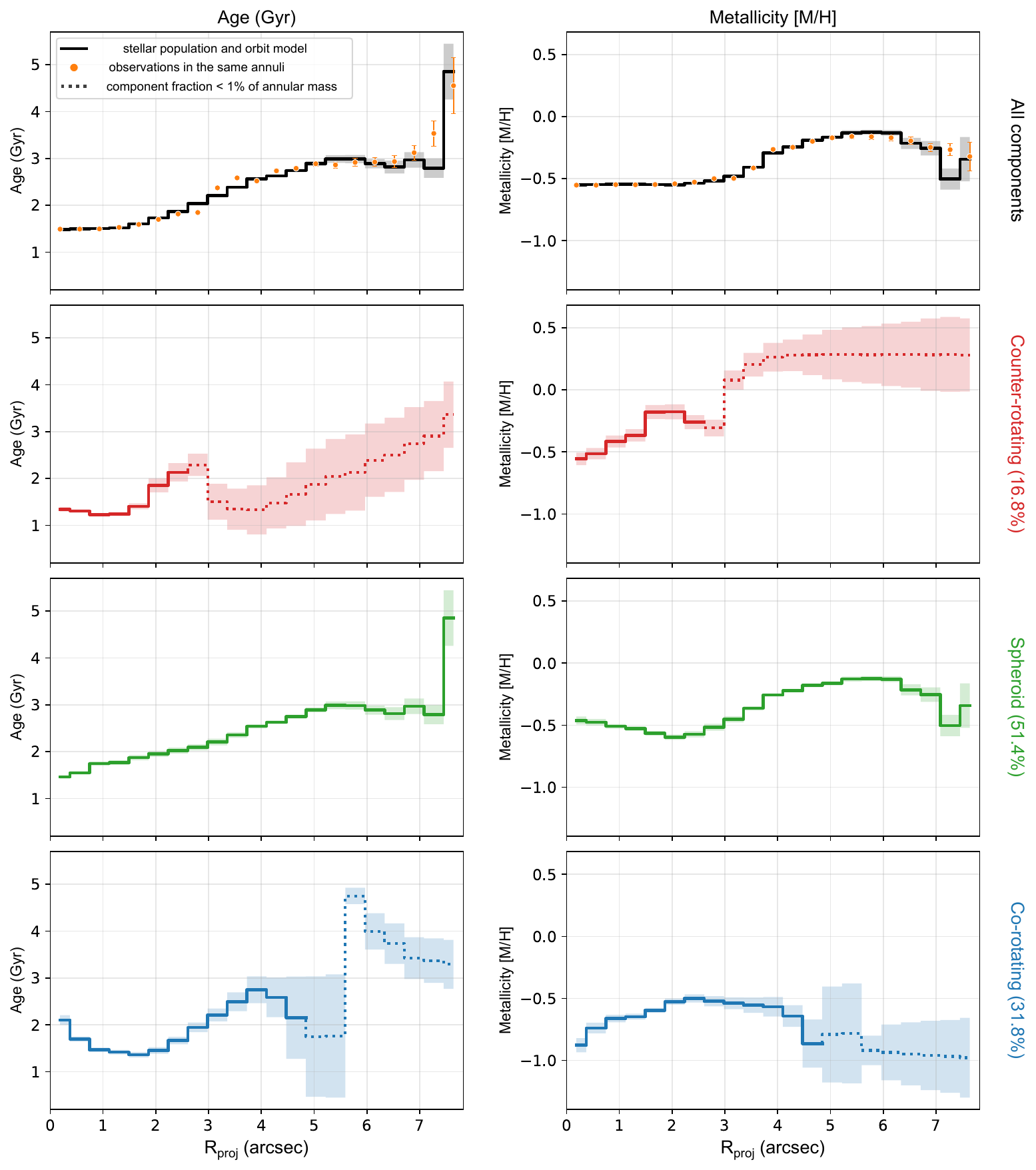}
\caption{Same as Fig.~\ref{ris:PGC_radprof}, but for
LEDA~2220522. Component colors: counter-rotating disk (red, $16.8\%$ of the mass in the field), spheroid (green, $51.4\%$), and co-rotating disk (blue, $31.8\%$). The outermost annulus, containing only $0.01\%$ of the total projected mass within the field, should be interpreted with caution.}
\label{ris:LEDA_radprof}
\end{figure*}

The radial profiles $T_\mathrm{SSP}(R_\mathrm{proj})$ and $[\mathrm{M/H}]_\mathrm{SSP}(R_\mathrm{proj})$, extracted for the three components using~\eqref{eq:radial_profile}, are shown in Figs.~\ref{ris:PGC_radprof} (PGC~35706) and \ref{ris:LEDA_radprof} (LEDA~2220522). The mean ages, metallicities, and $M/L$ values, weighted by the projected component masses within the \textsc{MaNGA} field, are summarized in Table~\ref{tab:agemet}.

A check of the age--metallicity correlation across radial annuli reveals no general anticorrelation of the profiles. For the total model, the Pearson coefficient is $r=+0.72$ for PGC~35706 and $+0.67$ for LEDA~2220522; for the observed profiles it is $+0.60$ and $+0.76$, respectively. A strong negative correlation is found only for the counter-rotating component of PGC~35706 ($r=-0.90$ in segments where the component fraction is at least $1\%$); for the co-rotating disks the coefficients are $-0.26$ and $-0.05$. These values are descriptive, because adjacent samples of the regularized profiles are not statistically independent. We do not report effective radii for the individual components: the limited \textsc{MaNGA} field allows only a half-mass radius within the observed aperture to be determined, which cannot be identified with the global $R_e$ of a component.

\begin{table}
\caption{Component ages, $[\mathrm{M/H}]$, and mass-to-light ratios $M/L$, weighted by their projected mass within the \textsc{MaNGA} field ($\pm\sigma_{\mathrm{MC}}$, $N=50$). In each realization, the $M/L$ map is recalculated from the perturbed age and metallicity maps using the \textsc{E-MILES} grid, after which the corresponding \textsc{BVLS} problem is solved independently.}
\label{tab:agemet}
\centering
\resizebox{\columnwidth}{!}{%
\begin{tabular}{llccc}
\hline
 & & co-rotating & spheroid & counter-rotating \\
\hline
\multirow{3}{*}{\shortstack[l]{PGC\\35706}}
 & age [Gyr] & $8.25\pm0.20$ & $8.66\pm0.09$ & $9.61\pm0.38$ \\
 & $[\mathrm{M/H}]$  & $+0.116\pm0.014$ & $+0.014\pm0.009$ & $-0.015\pm0.043$ \\
 & $M/L$  & $1.84\pm0.03$ & $1.92\pm0.01$ & $1.96\pm0.06$ \\
\hline
\multirow{3}{*}{\shortstack[l]{LEDA\\2220522}}
 & age [Gyr] & $1.61\pm0.06$ & $1.98\pm0.04$ & $1.32\pm0.04$ \\
 & $[\mathrm{M/H}]$  & $-0.613\pm0.024$ & $-0.465\pm0.016$ & $-0.450\pm0.043$ \\
 & $M/L$  & $0.39\pm0.02$ & $0.55\pm0.01$ & $0.34\pm0.02$ \\
\hline
\end{tabular}}
\end{table}

\subsubsection{PGC~35706}
\label{subsubsec:PGC}

The total PGC~35706 profile, weighted by projected mass in circular annuli (top row of Fig.~\ref{ris:PGC_radprof}), has an age of $T_\mathrm{SSP}\simeq9.3$~Gyr within the central $R_\mathrm{proj}<1\arcsec$. At radii of $1$--$4.5\arcsec$, the age remains within $8.4$--$8.8$~Gyr and then generally decreases, reaching $\simeq5.2$~Gyr in the outermost annulus. The metallicity is $[\mathrm{M/H}]\simeq+0.07$--$+0.10$ in the central region, approaches solar around $R_\mathrm{proj}\simeq3\arcsec$, and becomes subsolar at larger radii; the deepest local decrease, to $\simeq-0.19$, occurs at $R_\mathrm{proj}\simeq8.9\arcsec$. Thus, the previously obtained peak at $R\sim8$--$9\arcsec$ with $[\mathrm{M/H}]\simeq+0.35$ and the accompanying sharp age decrease are absent from the profile constructed directly from the projected mass within the field of view.

The spheroid, which contains $60.3\%$ of the mass within the field, dominates in every annulus. Its age decreases from approximately $9.5$~Gyr at the center to $\simeq6$~Gyr at $R_\mathrm{proj}\simeq7$--$9.5\arcsec$ and $4.8$~Gyr in the outermost annulus; its metallicity changes from near-solar at the center to $[\mathrm{M/H}]\simeq-0.24$ around $9.4\arcsec$. The co-rotating disk ($30.9\%$) has an age predominantly within $8$--$9.5$~Gyr out to $R_\mathrm{proj}\simeq8\arcsec$ and a higher central metallicity ($[\mathrm{M/H}]\simeq+0.23$), which approaches solar already at $R_\mathrm{proj}\simeq2\arcsec$. Its mean parameters are $\langle T\rangle_\mathrm{co}=8.25\pm0.20$~Gyr and $\langle[\mathrm{M/H}]\rangle_\mathrm{co}=+0.116\pm0.014$.

The counter-rotating disk contains only $8.8\%$ of the mass in the field and has a substantial projected contribution mainly at $R_\mathrm{proj}\lesssim5\arcsec$; at larger radii its fraction is usually below $1\%$ of the annular mass. Consequently, the corresponding dotted values, including segments at the upper age-grid bound of $14.1$~Gyr, should not be interpreted as a measured outer profile. The mean values over the entire observed projection are $\langle T\rangle_\mathrm{counter}=9.61\pm0.38$~Gyr and $\langle[\mathrm{M/H}]\rangle_\mathrm{counter}=-0.015\pm0.043$. Within the statistical MC analysis, the co-rotating disk is younger than the counter-rotating disk by $1.36\pm0.51$~Gyr and more metal-rich by $0.130\pm0.046$~dex (formally $2.7\sigma$ and $2.9\sigma$, respectively). These estimates do not include the systematic uncertainty associated with regularization, the dynamical model, and the single-population approximation; given the low mass of the counter-rotating component, the differences should be regarded as preliminary rather than as a reliable date for a separate accretion event.

\subsubsection{LEDA~2220522}
\label{subsubsec:LEDA}

The total LEDA~2220522 profile (top row of Fig.~\ref{ris:LEDA_radprof}) increases from approximately $1.5$~Gyr at the center to $\simeq3.0$~Gyr at $R_\mathrm{proj}\simeq5.5$--$7.3\arcsec$. The outermost value of $4.85$~Gyr refers to an annulus containing only $0.01\%$ of the projected mass in the field and does not characterize the galaxy as a whole. The metallicity remains near $[\mathrm{M/H}]\simeq-0.55$ within $R_\mathrm{proj}\simeq2.5\arcsec$, reaches $\simeq-0.12$ at $5.5$--$6\arcsec$, and then declines toward the edge. The previously described near-solar peak at $R\sim5$--$6\arcsec$ and decline to $[\mathrm{M/H}]\simeq-1$ at the edge are not reproduced.

The spheroid ($51.4\%$ of the mass in the field) becomes the dominant component at $R_\mathrm{proj}\gtrsim2\arcsec$ and almost entirely determines the total profile beyond $5\arcsec$. Its age increases from $\simeq1.5$~Gyr at the center to $\simeq3$~Gyr at $R_\mathrm{proj}\simeq5.5$--$7.3\arcsec$, and its mean parameters are $\langle T\rangle_\mathrm{sph}=1.98\pm0.04$~Gyr and $\langle[\mathrm{M/H}]\rangle_\mathrm{sph}=-0.465\pm0.016$. The co-rotating disk ($31.8\%$) has sufficient support from the projected data out to $R_\mathrm{proj}\simeq4.7\arcsec$: within this region, its age varies from approximately $1.4$ to $2.7$~Gyr, while its metallicity remains subsolar. Its mean values are $\langle T\rangle_\mathrm{co}=1.61\pm0.06$~Gyr and $\langle[\mathrm{M/H}]\rangle_\mathrm{co}=-0.613\pm0.024$. The dotted co-rotating disk values at larger radii, where its fraction is below $1\%$ of the annular mass, are not used to interpret the outer gradient.

The counter-rotating disk ($16.8\%$) is most prominent at the center, where its fraction in individual annuli reaches $20$--$41\%$, but decreases rapidly with radius; the last profile sample with a fraction above $1\%$ is at $R_\mathrm{proj}\simeq2.4\arcsec$. Within this region, its age is approximately $1.2$--$2.1$~Gyr. The mean values over the entire observed projection are $\langle T\rangle_\mathrm{counter}=1.32\pm0.04$~Gyr and $\langle[\mathrm{M/H}]\rangle_\mathrm{counter}=-0.450\pm0.043$.

Correcting the projected weighting reverses the sign of the age difference between the disks: the counter-rotating disk is younger than the co-rotating disk by $0.292\pm0.061$~Gyr. The disks also differ in metallicity: the counter-rotating component is more metal-rich by $0.163\pm0.057$~dex. The formal significance relative to input-map noise alone is $4.8\sigma$ and $2.8\sigma$, respectively; the sign of the age difference is preserved in all 50 MC realizations. However, with $N_\mathrm{MC}=50$, the distribution tail at this level is not directly resolved, and this estimate does not include systematic effects of regularization, the single-population assumption for a cell, or the choice of SSP and dynamical models. In addition, the moderate inclination of LEDA~2220522 ($i=27^\circ$) worsens the recovery of the age--circularity relation compared to an edge-on orientation \citep{Zhu2020}. We therefore regard the ordering of the mean values---the counter-rotating disk is younger and more metal-rich than the co-rotating disk---as the reliable result of the current model, without assigning physical meaning to the exact formal number of $\sigma$. The younger age has the sign expected for external formation of the secondary disk, whereas its higher rather than lower metallicity requires separate discussion in Section~\ref{sec:discussion}.

\subsection{Component mass-to-light ratios}
\label{subsec:res_ml}

The input $M/L$ map is calculated for each observational bin from its age and metallicity using the \textsc{E-MILES} grid, after which a separate linear problem is solved for this map with its own set of cell values $\phi_j^\Upsilon$ (Section~\ref{subsec:inverse}). The component $\langle M/L\rangle_C$ values in Table~\ref{tab:agemet} are obtained by averaging this independent \textsc{BVLS} solution with weights $M_j^\mathrm{FoV}$, rather than by substituting the mean $\langle T\rangle_C$ and $\langle Z\rangle_C$ into the nonlinear SSP relation. In PGC~35706, the values for the three components are similar: $1.84\pm0.03$ for the co-rotating disk, $1.92\pm0.01$ for the spheroid, and $1.96\pm0.06$ for the counter-rotating disk. Their ordering broadly follows the component age sequence, but the differences are small.

In the younger LEDA~2220522, the mass-to-light ratios are lower and likewise follow the corrected age ordering: the young counter-rotating disk has the lowest value ($0.34\pm0.02$), followed by the co-rotating disk ($0.39\pm0.02$), while the spheroid has the highest value ($0.55\pm0.01$). Thus, the $M/L$ of the counter-rotating disk is lower, rather than higher, than that of the co-rotating disk. At the same time, $M/L$ is not an independent age measurement, since its input map is constructed from the same $T_\mathrm{SSP}$ and $[\mathrm{M/H}]_\mathrm{SSP}$; a separate decomposition is needed to account correctly for the nonlinear SSP relation and explicitly associate population properties with orbital components. The absolute normalization inherits the systematic uncertainties of the adopted SSP grid, IMF, and photometric calibration.

\begin{figure*}[t!]
\includegraphics[width=\linewidth]{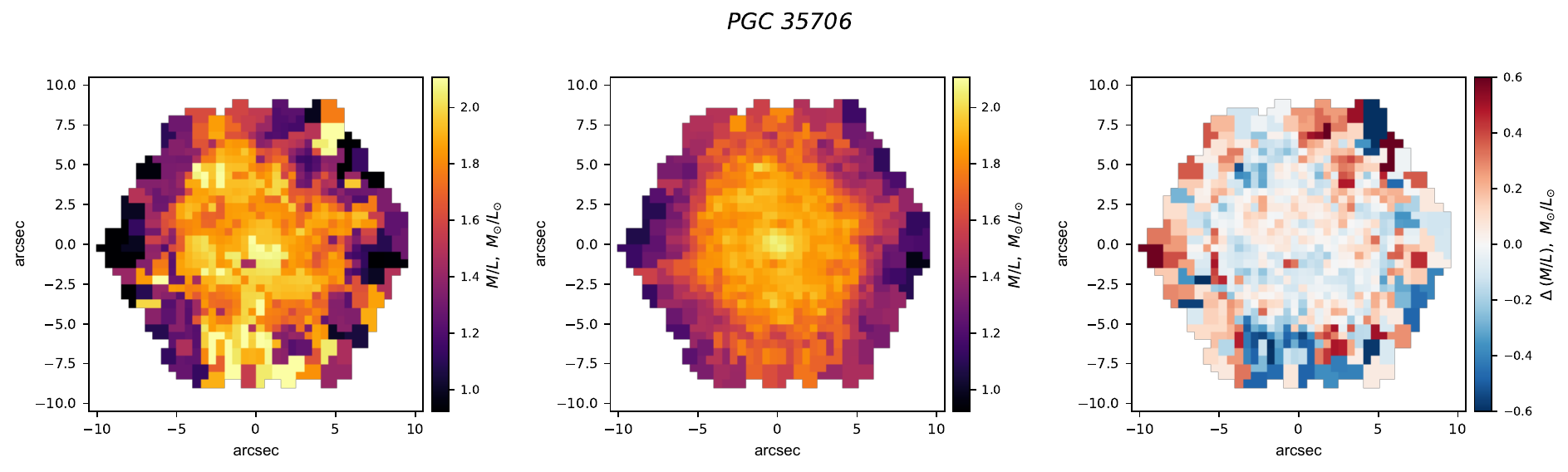}
\caption{Mass-to-light ratio $M/L$ for PGC~35706 at two stages of the stellar population and orbit model, and their difference. \textit{Left:} population $M/L$ by spaxel (\textsc{E-MILES}, from age and metallicity); \textit{center:} orbital $M/L$ from the \textsc{BVLS} assignment to orbits; \textit{right:} their difference (orbital $-$ population). All maps are in population $M/L$ units; when determining the global normalization of the dynamical model, the velocities are additionally scaled by $\sqrt{\Upsilon}$ (physical $M/L\approx1.3$, Section~\ref{subsec:data_halo}).}
\label{ris:ml_stages_PGC}
\end{figure*}

\begin{figure*}[t!]
\includegraphics[width=\linewidth]{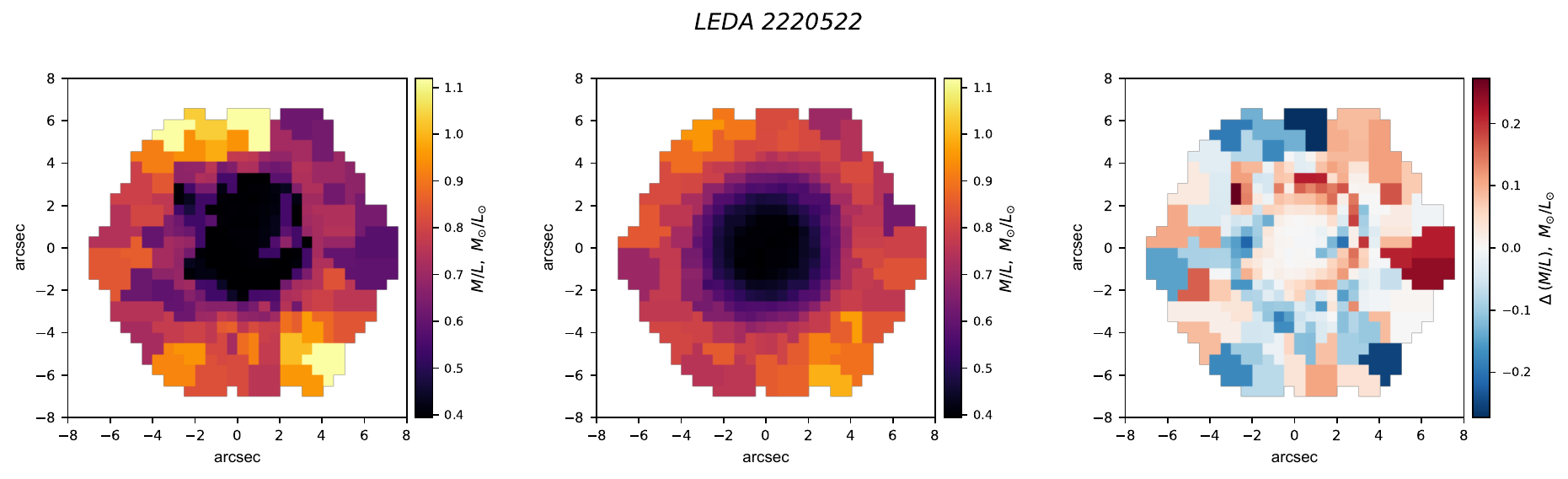}
\caption{Same as Fig.~\ref{ris:ml_stages_PGC}, but for LEDA~2220522. The radial increase in $M/L$ from the young center ($\approx0.4$) to the older outskirts ($\approx0.9$) is visible at both stages.}
\label{ris:ml_stages_LEDA}
\end{figure*}

The progression of $M/L$ through the model stages and its spatial distribution are shown in Figs.~\ref{ris:ml_stages_PGC} (PGC~35706) and \ref{ris:ml_stages_LEDA} (LEDA~2220522): the population map by spaxel (\textsc{E-MILES}, from age and metallicity), the orbital $M/L$ from the \textsc{BVLS} assignment, and their difference. In LEDA~2220522, both stages show a radial increase in $M/L$ from the young center ($\approx0.4$) to the older outskirts ($\approx0.9$); in PGC~35706, whose components are all old, the ratio is high throughout the field, with a shallow central maximum.

\subsection{Monte Carlo stability of the solution}
\label{subsec:res_mc}

\begin{figure*}[ht!]
\includegraphics[width=\linewidth]{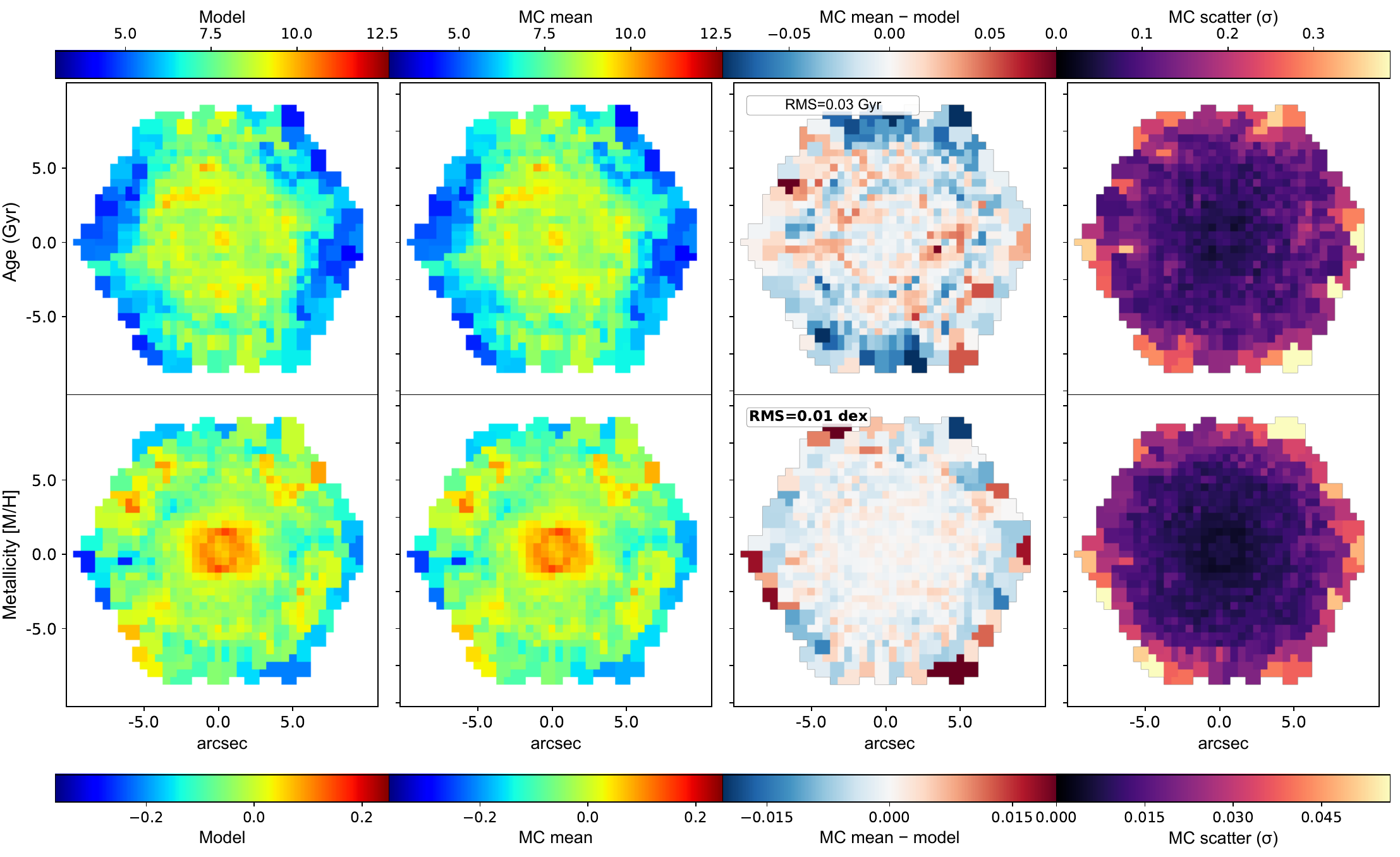}
\caption{Monte Carlo assessment of the stability of the stellar population and orbit
solution for PGC~35706. \textit{From left to right:} model map (original solution), mean over $N_\mathrm{MC}=50$ realizations, their difference ``MC mean $-$ model'' (a measure of stability; on a diverging scale symmetric about zero, with its RMS in the corner), and standard deviation over the MC ensemble (a separate column scale starting at zero). Top row: age; bottom row: metallicity.}
\label{ris:PGC_mc_stab}
\end{figure*}

\begin{figure*}[ht!]
\includegraphics[width=\linewidth]{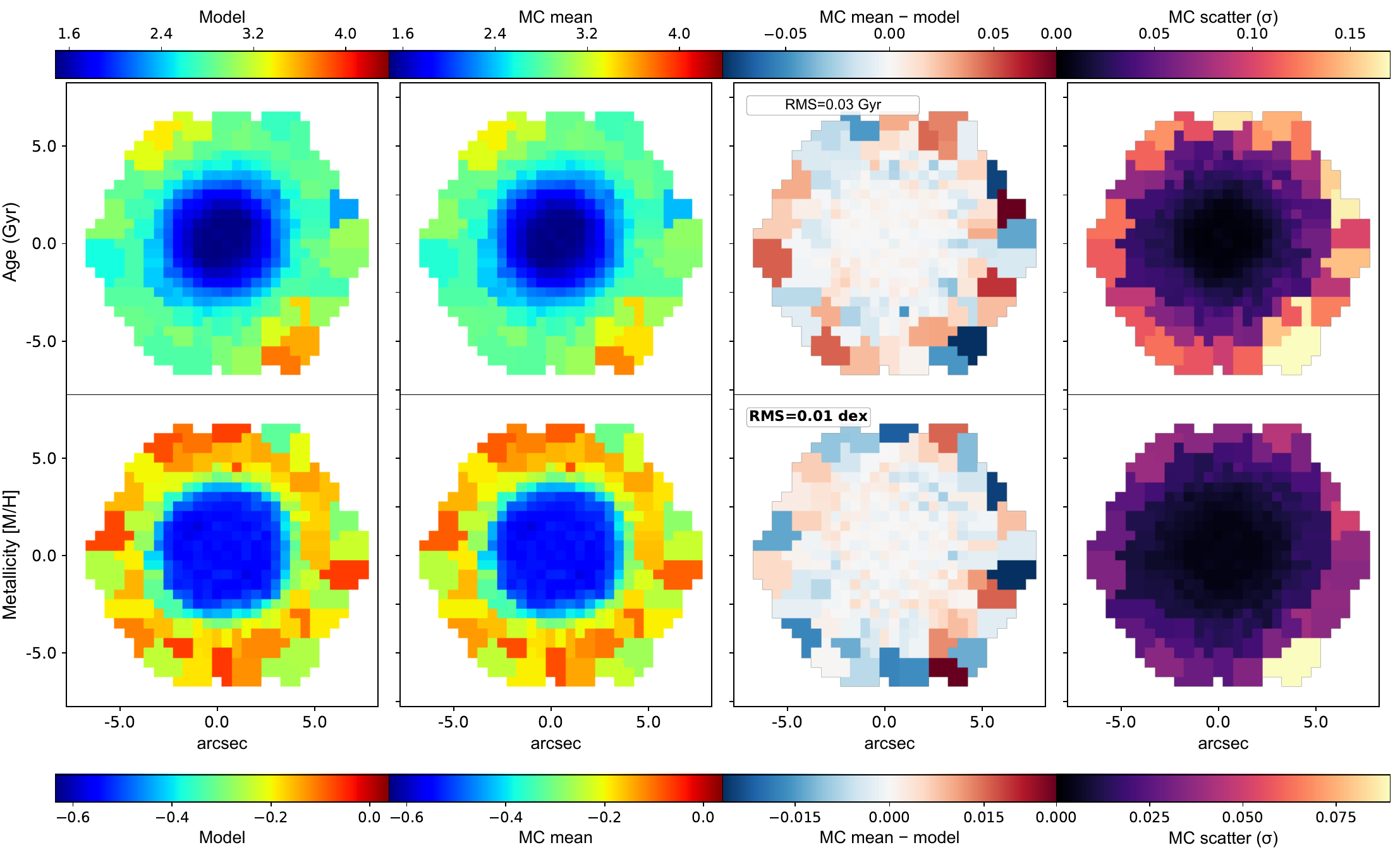}
\caption{Same as Fig.~\ref{ris:PGC_mc_stab}, but for
LEDA~2220522.}
\label{ris:LEDA_mc_stab}
\end{figure*}

The results of the Monte Carlo analysis of the two-dimensional maps are shown in Figs.~\ref{ris:PGC_mc_stab} and \ref{ris:LEDA_mc_stab}. The RMS difference between the original projected map and the MC mean is $0.043$~Gyr in age and $0.0056$~dex in metallicity for PGC~35706, and $0.015$~Gyr and $0.0080$~dex, respectively, for LEDA~2220522. Thus, the observed projections of the solution are stable against statistical noise in the input maps. This test does not, however, establish the uniqueness of the set of values $\phi_j$ in individual phase-space cells: because of the effective rank deficiency, different cell solutions may produce similar projected maps, while regularization and active bounds further narrow the MC scatter.

For PGC~35706, the median $\sigma_\mathrm{MC}$ of the projected map is $0.113$~Gyr in age and $0.014$~dex in metallicity; the $98$th percentiles are $0.436$~Gyr and $0.054$~dex. For LEDA~2220522, the corresponding medians are $0.061$~Gyr and $0.019$~dex, and the $98$th percentiles are $0.157$~Gyr and $0.085$~dex. The maximum values over the entire field do not exceed $0.53$~Gyr and $0.09$~dex. These quantities characterize the stability specifically of the projection onto the plane of the sky for a fixed dynamical model, grid, and regularization.

The $\pm1\sigma_\mathrm{MC}$ bands in Figs.~\ref{ris:PGC_radprof}, \ref{ris:LEDA_radprof} depend primarily on the projected contribution of a component in a particular annulus, rather than solely on its total fraction in the field. In the PGC~35706 segments where the component fraction is at least $1\%$, the median $\sigma_\mathrm{MC}$ of the counter-rotating disk profile is $0.40$~Gyr and $0.050$~dex, and the maximum is $0.68$~Gyr and $0.24$~dex. These values are not several times larger than those of the more massive components; conditioning deteriorates sharply where the local fraction becomes negligibly small, as marked by the dotted lines. Thus, the global mass fraction alone does not define a universal threshold for the applicability of the method, and formal component uncertainties must be interpreted together with the local mass fraction and the systematic limitations of the model.

\section{DISCUSSION}
\label{sec:discussion}

\subsection{LEDA~2220522: a young and chemically enriched
counter-rotating disk}
\label{subsec:disc_LEDA}

The corrected stellar population and orbit decomposition of LEDA~2220522 identifies a younger counter-rotating disk: $\langle T\rangle_\mathrm{counter}=1.32\pm0.04$~Gyr versus $1.61\pm0.06$~Gyr for the co-rotating disk, with mass fractions of $16.8\%$ and $31.8\%$, respectively. The sign of the age difference agrees with the classical picture of secondary-disk formation from retrograde gas supplied later, proposed for NGC~5719 \citep{Coccato2011}, NGC~3593 and NGC~4550 \citep{Coccato2013, Johnston2013}, NGC~1366 \citep{Morelli2017}, and NGC~4191 \citep{Coccato2015}. The chemical part of this picture, however, does not hold: the counter-rotating disk is more metal-rich, rather than more metal-poor, with $\langle[\mathrm{M/H}]\rangle_\mathrm{counter}=-0.450\pm0.043$ versus $-0.613\pm0.024$ for the co-rotating disk.

An external origin of the counter-rotating disk is independently supported by the gas kinematics. The kinematic major axis of the gas field is close in orientation to the stellar one, but the line-of-sight velocity gradient has the opposite sign to that of the light-dominant stellar component (the correlation coefficient between the gaseous and stellar line-of-sight velocity fields is $\approx-0.9$). Thus, the gas rotates in the same direction as the counter-rotating stellar disk. This correspondence, already noted by \citet{Goradzhanov2025}, links the observed gas to the counter-rotating component. The gas ionization is consistent with a star formation mechanism: on the Baldwin--Phillips--Terlevich diagnostic diagram (BPT; \citealt{Baldwin1981}), it lies in the star-forming galaxy region rather than the AGN/LINER region. The gas metallicity from the O3N2 and N2 calibrations \citep{Marino2013} is near-solar: $12+\log(\mathrm{O/H})\approx8.6$ ($8.63$ and $8.58$, respectively), or $[\mathrm{O/H}]\approx-0.1$ for a solar value of $8.69$ \citep{Asplund2009}, with an almost flat radial profile.

The minimal interpretation of these data does not require two alternating accretion episodes with opposite angular momentum. It is sufficient for the co-rotating disk to have existed before the supply of retrograde gas from which the younger counter-rotating component subsequently formed; the remaining gas continues to rotate in the same direction and to participate in current star formation. The higher metallicity of the counter-rotating stars and the near-solar oxygen abundance of the gas show that the accreted material was not pristine. Possible sources remain chemically enriched gas from a gas-rich satellite or previously processed gas from the intragroup or circumgalactic medium. The offset of LEDA~2220522 from its group by $\approx2\sigma_v$ and the presence of a satellite at a projected separation of $\approx48$~kpc make an external source plausible, but do not distinguish between these possibilities.

Long-lived counter-rotating disks form both through retrograde gas inflow \citep{Thakar1996, Algorry2014, Starkenburg2019, Khoperskov2021} and through a minor merger with a gas-rich satellite \citep{Bassett2017}; the mass of LEDA~2220522, $M_*\sim1.0\times10^{10}$~\Ms\ from the \textsc{NSA} catalog \citep{Blanton2011}, does not allow a choice between these formation scenarios. The available data also do not yield a precise onset time for accretion: SSP-equivalent ages characterize averaged populations rather than directly specifying the times of dynamical events. The formal differences of $0.292\pm0.061$~Gyr and $0.163\pm0.057$~dex account only for input-map noise; the comparison of gaseous $[\mathrm{O/H}]$ with stellar $[\mathrm{M/H}]$ also remains qualitative because the scales differ. A more rigorous chronology will require an independent chronometer, such as $\alpha$-element abundances, and data capable of distinguishing prolonged inflow from a minor merger.

\subsection{PGC~35706: preliminary population differences in a
low-mass counter-rotating disk}
\label{subsec:disc_PGC}

The counter-rotating disk of PGC~35706 contains only $8.8\%$ of the projected mass within the \textsc{MaNGA} field, but after correct weighting its mean parameters differ from those of the co-rotating disk. Within the statistical MC analysis, the counter-rotating component is older by $1.36\pm0.51$~Gyr and more metal-poor by $0.130\pm0.046$~dex, corresponding to formal differences of $2.7\sigma$ and $2.9\sigma$. Thus, the previous conclusion that the two disks are entirely indistinguishable is not supported by the statistical errors of the input maps.

The formal precision of the mean values does not imply that the spatial structure of the low-mass component has been reconstructed over the entire field. Its contribution exceeds $1\%$ of the annular mass mainly only at $R_\mathrm{proj}\lesssim5\arcsec$; the outer dotted samples, including values at the age-grid bound, are effectively unconstrained by the observations. Moreover, the reported significance does not include changes in the result due to choosing different regularization, a different dynamical model or SSP grid, or relaxing the single-population approximation. The detected differences should therefore be regarded as evidence requiring verification, rather than as precise dating and chemical characterization of a separate accretion event.

The kinematic identification of the counter-rotating component remains reliable, and the agreement between its rotation direction and that of the \Ha\ gas supports an external origin of the retrograde material. The rich environment of PGC~35706---its position as the second most luminous member of a group of about 11 galaxies, a brighter central galaxy, and nearby satellites---also makes an accretion scenario plausible. At the same time, an older and more metal-poor counter-rotating component could reflect either an early supply of external material or a biased result caused by the small projected component fraction and the sensitivity of the decomposition to regularization and the dynamical model (Section~\ref{subsec:disc_limits}); the available data cannot distinguish these possibilities. Verification will require spectroscopy at higher spatial resolution and repetition of the analysis with alternative dynamical models and regularization parameters.

\subsection{Method limitations and systematic effects}
\label{subsec:disc_limits}

The analysis presented here relies on several approximations, each of which introduces systematic effects not included in the MC uncertainty $\sigma_\mathrm{MC}$.

\emph{The single stellar population approximation within an orbital cell.} Each $(R,\lambda_z)$ cell in the model has a single pair $(\phi^T_j,\phi^Z_j)$, whereas real dynamical subsystems may contain a mixture of stellar populations with different ages and metallicities. This limitation has already been emphasized in the literature \citep{Poci2019, Zhu2020, Jin2024} and in some cases is overcome by moving to a BVLS problem with full spectral templates for each cell (the full spectral fitting version of the method). We do not make this transition here, because the limited spectral resolution of \textsc{MaNGA} and the relatively small field of view do not allow a detailed star formation history to be extracted from each individual cell. The systematic effect of the single-population assumption is most evident in the central region of PGC~35706 (Fig.~\ref{ris:PGC_residuals}), where the observed $[\mathrm{M/H}]$ gradient exceeds the spatial scale of the adopted $(R,\lambda_z)$ grid.

\emph{The axisymmetric dynamical model.} The dynamical framework is constructed in the axisymmetric approximation (Section~\ref{subsec:data_dyn_input}). Possible departures of the potential from axial symmetry (triaxiality, a bar) may change the projected orbital weights and thereby transfer signal between $\lambda_z$ regions. This limitation is particularly important for PGC~35706, because its gas field shows substantial noncircular motions: the RMS residual of a simple rotating-disk model is $41$~km/s, compared to a median formal error of $6$~km/s, and the two sides of the rotation curve differ by an average of $25$~km/s (Section~\ref{subsec:data_halo}). Good agreement between the model and observed stellar kinematic maps rules out a gross mismatch between the model and data, but does not by itself quantify this systematic bias in the population decomposition.

\emph{Accuracy of the input stellar population maps.} The errors in $T_\mathrm{SSP}$ and $[\mathrm{M/H}]_\mathrm{SSP}$ in each Voronoi bin, provided by \nb, are treated as independent in the MC procedure. Our implementation does not account for the actual covariance between these parameters, or their correlation with the coefficients of the polynomial continuum fit and with $\alpha$-element abundance. Extending the method by simultaneously extracting the $[\mathrm{Mg/Fe}]$ index from \textsc{E-MILES} models with variable $\alpha$ abundances \citep{Vazdekis2015} and independently calibrating the duration of star formation using $[\alpha/\mathrm{Fe}]$ \citep{Thomas2005, delaRosa2011} appears to be a promising direction for future work, which would allow formation scenarios for counter-rotating disks to be distinguished more rigorously.

\emph{Regularization and active bounds.} The choice of $\lambda=0.01$ is based on a heuristic knee in $\chi^2_\mathrm{red}(\lambda)$, rather than on a probabilistic model of a prior distribution. Even at the adopted $\lambda$, $30$--$39\%$ of cell values remain at the bounds of the allowed range, depending on the galaxy and parameter. The MC procedure repeats the calculation with fixed $L$, $\lambda$, and bounds and therefore does not include uncertainty associated with the choice of these settings. Stability of the projected maps should not be equated with uniqueness of the solution in the $(R,\lambda_z)$ plane.

\emph{Limited field of view and support from projected data.} All mass fractions, component means, and profiles refer only to the actual \textsc{MaNGA} field. Circular annuli at large radii are only partly filled because of the hexagonal field shape, and the fraction of an individual component may vanish well before the edge of the observations. The dotted profile segments and the outermost low-mass annulus of LEDA~2220522 are therefore not used for physical interpretation, and the inferred fractions cannot be extrapolated to the entire galaxy without an additional model.

\emph{Uncertainty in the dynamical model and dark halo.} For LEDA~2220522, the halo normalization within the \textsc{MaNGA} field is poorly constrained: a broad range of $V_{c,\mathrm{DM}}(R_e)\simeq35$--$90$~km/s and $f_\mathrm{DM}(<R_e)\simeq10$--$56\%$ is statistically allowed. PGC~35706 constrains the central dark mass much better, with a preferred $f_\mathrm{DM}(<R_e)\simeq6\%$ (Section~\ref{subsec:data_halo}). The current MC analysis fixes one dynamical model and does not propagate the uncertainty across the entire allowed LEDA plateau to the orbital populations. Agreement between solutions with isothermal and \textsc{NFW} halos shows the absence of a strong dependence on the adopted profile, but does not eliminate this systematic uncertainty.

Given these limitations, the most stable result for LEDA~2220522 within the fixed model is the ordering of the mean parameters: the counter-rotating disk is younger and more metal-rich than the co-rotating disk, and the \Ha\ gas rotates in the same direction as the young component. Together, these signatures support the late supply of chemically enriched retrograde gas, but do not prove a specific source of the material and do not require two separate accretion episodes with opposite directions. For PGC~35706, the differences in mean parameters have moderate formal significance, but the small local contribution of the counter-rotating disk and the systematic uncertainties of the decomposition do not yet allow them to be translated into a reliable chronology.

\section{CONCLUSIONS}
\label{sec:conclusions}

This work presents the results of stellar population and orbit modeling of two galaxies with kinematically misaligned components---PGC~35706 and LEDA~2220522---as a direct continuation of the Schwarzschild dynamical model we previously constructed in \citet{Goradzhanov2025}. The fundamentally new step is to assign the $T_\mathrm{SSP}$, $[\mathrm{M/H}]_\mathrm{SSP}$, and $M/L$ maps to orbital cells in the $(R,\lambda_z)$ plane through a regularized bounded inverse problem, and then to construct profiles using projected component mass directly within the observed \textsc{MaNGA} field. This allows us to move from purely kinematic component identification to comparison of their stellar populations. The main results can be stated as follows:

\begin{enumerate}
\item \textbf{Structural component mass fractions.} In PGC~35706, the spheroid contains $60.3\%$ of the projected mass within the \textsc{MaNGA} field, the co-rotating disk $30.9\%$, and the counter-rotating disk $8.8\%$. In LEDA~2220522, the corresponding fractions are $51.4\%$, $31.8\%$, and $16.8\%$. These quantities refer only to the observed field and are not estimates of the total component masses throughout the galaxy.

\item \textbf{Reproduction of the observed maps.} The model reproduces the structure of the observed age maps in both galaxies; the RMS residual is $\sim\!0.6$~Gyr for LEDA~2220522 and $\sim\!2$~Gyr for the older PGC~35706, where it is determined mainly by noise in the observed map at the edge of the field of view. The metallicity maps are reproduced satisfactorily (residual $\sim\!0.12$~dex). No large-scale coherent residual is found in the central region of PGC~35706; the most prominent local discrepancy is confined to two adjacent bins with $\Delta Z=-0.22$ and $-0.24$~dex.

\item \textbf{LEDA~2220522: a young enriched counter-rotating disk.} The counter-rotating disk has an age of $1.32\pm0.04$~Gyr and is younger than the co-rotating disk ($1.61\pm0.06$~Gyr) by $0.292\pm0.061$~Gyr. It is also more metal-rich: $[\mathrm{M/H}]_\mathrm{counter}=-0.450\pm0.043$ versus $-0.613\pm0.024$, a difference of $0.163\pm0.057$~dex. Its lower $M/L=0.34\pm0.02$, compared to $0.39\pm0.02$ for the co-rotating disk, is consistent with this age ordering. The agreement between the rotation directions of the \Ha\ gas and the young counter-rotating component supports a scenario involving the late supply of chemically enriched retrograde gas, but does not require two separate episodes with opposite angular momentum or uniquely identify the source of the material.

\item \textbf{PGC~35706: preliminary differences in the low-mass component.} The counter-rotating disk is older than the co-rotating disk by $1.36\pm0.51$~Gyr and more metal-poor by $0.130\pm0.046$~dex, corresponding to moderate formal significance relative to input-map noise. This combination differs from the commonly observed picture of a young secondary component and, if confirmed with alternative dynamical models, would make PGC~35706 an unusual case among galaxies with counter-rotating disks. However, the counter-rotating component contains only $8.8\%$ of the mass in the field and has sufficient local support mainly at $R_\mathrm{proj}\lesssim5\arcsec$. The differences between the means should therefore be considered preliminary until tested with alternative dynamical models, regularization parameters, and data at higher spatial resolution.

\item \textbf{Stability and applicability.} The MC analysis with $N_\mathrm{MC}=50$ confirms the stability of the projected maps against statistical noise: the RMS difference between the nominal map and the MC mean does not exceed $0.043$~Gyr in age or $0.008$~dex in metallicity. This does not prove uniqueness of the solution in individual phase-space cells. The reliability of a component profile is determined by its local projected fraction within an annulus, rather than by a universal threshold in the total component mass; segments with fractions below $1\%$ are marked with dotted lines and are not used for physical interpretation. The main systematic uncertainties are associated with regularization, the single-population assumption for a cell, the dynamical model, the SSP grid, and the limited field of view.
\end{enumerate}

The results demonstrate that the stellar population and orbit approach, previously applied to \textsc{MUSE} \citep{Poci2019, Poci2021, Zhu2022a, Zhu2022b, Ding2023} and \textsc{CALIFA} \citep{Jin2024} data, is also informative for \textsc{MaNGA} data when the projected contribution of the identified component is explicitly taken into account as a function of radius. A universal applicability threshold based solely on the total mass fraction cannot be established from the two objects considered here. Promising directions include adaptive partitioning of the phase-space grid, propagation of dynamical model and regularization uncertainties, and accounting for $\alpha$-element abundances \citep{Thomas2005, delaRosa2011} as an independent constraint on the duration of star formation and the scenario for the supply of external gas.

\section*{ACKNOWLEDGMENTS}

This research was carried out within the framework of the state assignment of M.~V.~Lomonosov Moscow State University.